\documentclass[12pt]{revtex4-2}

\usepackage{natbib}
\usepackage{graphicx}
\usepackage{amsfonts}
\usepackage{amsmath,amssymb} 
\usepackage{bm}
\usepackage{xcolor}
\usepackage{hyperref}
\usepackage{caption}
\usepackage{subcaption}
\usepackage{tabularx}
\usepackage{ulem} 
\usepackage{mathtools}

\input{CQGformat.input}
\newcommand{\En}{\mathcal{E}}
\newcommand{\Lz}{\mathcal{L}}
\newcommand{\Q}{\mathcal{Q}}
\newcommand{\K}{\mathcal{K}}
\newcommand{\sn}{\mathrm{sn}}
\renewcommand{\sp}{\epsilon} 
\newcommand{\mr}{\epsilon} 
\newcommand{\Ad}{{\rm{Ad}}} 
\newcommand{\PA}{{\rm{SF}}} 

\newcommand{\diss}{{\rm{diss}}}
\newcommand{\cons}{{\rm{cons}}}

\newcommand{\avg}[1]{\left\langle #1 \right\rangle}
\newcommand{\bavg}[1]{\big\langle #1 \big\rangle}
\newcommand{\osc}[1]{\Breve{#1}}
\newcommand{\nit}[1]{\Tilde{#1}}
\newcommand{\PD}[2]{\frac{\partial #1}{\partial #2}}
\newcommand{\HOT}[1]{\mathcal{O}( \sp ^{ #1 } )} 

\begin{document}

\title{Eccentric self-forced inspirals into a rotating black hole}

\author{Philip Lynch$^1$, Maarten van de Meent$^2$, Niels Warburton$^1$}

\address{$^1$ School of Mathematics and Statistics , University College Dublin, Belfield, Dublin 4}

\address{$^2$ Max Plank Institute for Gravitational Physics (Albert Einstein Institute), Potsdam-Golm, Germany}

\ead{philip.lynch@ucdconnect.ie}

\begin{abstract}

	We develop the first model for extreme mass-ratio inspirals (EMRIs) into a rotating massive black hole driven by the gravitational self-force.
	Our model is based on an action angle formulation of the method of osculating geodesics for eccentric, equatorial (i.e., spin-aligned) motion in Kerr spacetime.
	The forcing terms are provided by an efficient spectral interpolation of the first-order gravitational self-force in the outgoing radiation gauge.
	We apply a near-identity (averaging) transformation to eliminate all dependence of the orbital phases from the equations of motion, while maintaining all secular effects of the first-order gravitational self-force at post-adiabatic order. 
	This implies that the model can be evolved without having to resolve all $\mathcal{O}(10^5)$ orbit cycles of an EMRI, yielding an inspiral model that can be evaluated in less than a second for any mass-ratio.
	In the case of a non-rotating central black hole, we compare inspirals evolved using self-force data computed in the Lorenz and radiation gauges. 
	We find that the two gauges generally produce differing inspirals with a deviation of comparable magnitude to the conservative self-force correction. 
	This emphasizes the need for including the (currently unknown) dissipative second order self-force to obtain gauge independent, post-adiabatic waveforms.
\end{abstract}
\noindent{\it Extreme Mass Ratio Inspirals, Gravitational Self-Force, Celestial Mechanics\/}


\maketitle

\section{Introduction}
The detection of the first gravitational wave signal \cite{LIGOScientific:2016aoc} ushered in a new era of astronomy, with ground-based observatories having now observed just over a hundred signals \cite{LIGOScientific:2018mvr, LIGOScientific:2020ibl, LIGOScientific:Cat3}. 
The next generation of space-based detectors, such as the Laser Interferometer Space Antenna (LISA), will probe the previously inaccessible millihertz band of the gravitational wave spectrum, allowing for the detection of hitherto unseen gravitational wave sources. 
Among these sources are extreme mass-ratio inspirals (EMRIs), which consist of a stellar mass compact object (such as a black hole or neutron star) spiralling into a supermassive black hole. 
These systems are characterised by their extremely small mass ratio, typically between $10^{-4}$ and $ 10^{-7}$. 
Unlike the signals detected by ground-based detectors, EMRIs will radiate in the LISA frequency band for up to hundreds of thousands of orbital cycles \cite{Babak2017a}. 
They are also expected to be eccentric and precessing, potentially resulting in multi-year long waveforms with rich and complex morphology \cite{Drasco2014}.
These signals encode the spacetimes of supermassive black holes, promising exquisite parameter estimation and some of the most sensitive probes for new physics beyond general relativity \cite{Berry2019}. 

The majority of EMRIs  will have a very low instantaneous signal-to-noise ratio (SNR), and so the data must be processed with matched filtering techniques which will allow for the build up of the SNR over time \cite{Barack2004}. 
Such techniques require the development of theoretical waveform templates to compare against the data. 
To achieve LISA's science objectives, these templates need their phase to be accurate to within a fraction of a radian, even after hundreds of thousands of orbital cycles. 
They also need to be extensive across the large parameter space of possible EMRI configurations. 
Moreover, since many template evaluations would be needed, they should also be fast to compute, ideally in less than a second.

To meet this challenge, several so-called ``kludge'' models have been developed, which are both extensive and quick to compute \cite{Kennefick2002,Babak2007,Sopuerta2011,Chua2015}. 
However, they also make use of non-relativistic approximations which cause them to fall short of the accuracy requirement, though they may still be sufficient for the detection of loud EMRI signals \cite{Chua2017}. 
Despite their shortcomings, these models are invaluable for testing data analysis techniques for LISA through the mock data challenges \cite{Arnaud2006, Babak2009,Chua:2021aah}.
In order to detect more EMRI signals, relativistic, ``adiabatic" waveforms are required. The adiabatic trajectory can be calculated by balancing the fluxes of energy and angular momentum through null infinity the event horizon with the energy and angular momentum lost by the secondary throughout the inspiral.
This has been implemented in a practical framework for non-spinning, eccentric EMRIs \cite{Chua2021a,Katz2021}.
While this represents a significant accuracy improvement over the kludge models, work remains to be done to extend this framework to generic inspirals into rotating black holes.

To detect all EMRI signals and enable precision parameter estimation, requires ``post-adiabatic" waveforms.
Producing such waveforms requires calculating the local force experienced by the secondary due to the presence of its own gravitational field, known as the gravitational self-force (GSF).
This can be calculated perturbatively by expanding the field equations in powers of the small mass ratio of the binary, which makes this approach ideally suited to EMRIs.
Post-adiabatic EMRI waveforms will require not only full knowledge of the the first-order self-force, but also the orbit averaged contribution of the second-order self-force \cite{Hinderer2008, VanDeMeent2020a}.

At each instant, the self-force is a functional of the the past history of the secondary which can make it challenging to compute. 
One approach is to couple the field equations and the equations of motion and self-consistently solve both in a time-domain simulation.
While this has been implemented for a toy model of a particle carrying a scalar charge orbiting a Schwarzschild black hole \cite{Diener2012}, numerical stability issues have so far stifled similar attempts for the gravitational case \cite{Dolan2013}. 
Moreover, this approach is computationally very expensive, making it impractical for generating large numbers of templates. 
However, it does promise waveforms against which more efficient schemes should be tested.

An alternative method is to compute the self-force for a body moving along fixed geodesics of the background spacetime and then use that force to move to another geodesic at a later timestep.
The periodic nature of these geodesic orbits allows for calculations in frequency domain leading to many efficient calculations of the first-order self-force in both Schwarzschild \cite{Barack2007,Barack2010b,Akcay2013,Osburn2014} and Kerr \cite{VandeMeent2016,VanDeMeent2018} spacetimes. 
Second-order calculations are also emerging, though at present these are restricted to quasi-circular inspirals into non-rotating black holes \cite{Pound2019,Warburton2021}.
These calculations can be repeated across the parameter space of bound geodesics and then interpolated in a preprocessing step, as has been done for eccentric inspirals into a Schwarzschild black hole \cite{Warburton2012,Osburn2016}. 
One of the goals of this work is to compute self-forced inspirals into rotating (Kerr) black holes.

With an interpolated model of the geodesic self-force in hand, one can restate the EMRI equations of motion in a more convenient form for numerical integration using the method of ``osculating orbital elements" (or ``osculating geodesics" when applied to the relativistic context). 
In this approach, the inspiral is described as a smooth evolution through neighbouring geodesics that are instantaneously tangent to the true inspiral. 
Formally, one identifies a set of constants of motion which uniquely identify a geodesic, known as ``orbital elements". 
These constants are then promoted to functions of time which are governed by a set coupled ordinary differential equations that are derived from the ``osculating conditions".
These new equations of motion are then solved numerically to obtain the inspiral trajectory of the secondary.
There are a number of equivalent formulations for these ODEs which have been derived for both Schwarzschild \cite{Pound2008} and Kerr \cite{Gair2011,Pound2021} inspirals. 
In this work, we make use of a formulation based on action angles of the geodesic motion that was sketched out in Ref.~\cite{Gair2011}.

However, for all of these formulations, the resulting equations of motion depend on the orbital phases. 
This means that the numerical integrator will have to resolve features on the orbital timescale, requiring the use of many small time steps.
Since a typical EMRI undergoes $\sim10^4 - 10^6$ orbital cycles during a radiation reaction timescale, this results in computational times of minutes to hours for a single inspiral, depending on the orbital configuration. 

Following Ref.~\cite{NITs} (hereafter Paper I), we overcome this problem by applying a near-identity (averaging) transformation (NIT) \cite{Kevorkian1987} to the self-forced equations of motion. 
These transformed equations have two important properties:  (i) they no longer depend on the orbital phase, and (ii) they capture the long-term secular evolution of the original inspiral to the same order of approximation in the mass ratio as the original equations of motion.
The first property means the transformed equation of motion can be numerically solved for any mass ratio in less than a second as the numerical integrator no longer needs to resolve the thousands of oscillations on the orbital timescale.

This approach has been applied to the case of eccentric inspirals into non-rotating black holes \cite{NITs,McCart2021}.
In this work we apply the NITs to orbital motion in the Kerr spacetime.
Combined with an interpolated model of first-order gravitational self-force data, these averaged equations of motion allow us to efficiently compute inspirals around a Kerr black hole which, for the first time, include all first order in the mass ratio effects. 
Since these inspirals are fast to compute, this approach can provide EMRI waveforms useful for practical data analysis when coupled to a fast waveform generation scheme, e.g., \cite{Chua2021a}.

At this point, we emphasize that the inspirals we present do not reach the sub-radian accuracy required for EMRI data analysis as there are not yet any second order self-force calculations in this domain.
To stress the importance of the second order contribution we examine the effects of driving the inspirals using first-order self-forces computed in two different gauges and demonstrate explicitly that without the inclusion of the second-order self-force the inspiral phase is not gauge invariant.
Nonetheless, the framework we present can readily incorporate new self-force results as they become available.

This paper is laid out as follows.
In Sec.~\ref{section:EoM}, we review geodesic motion and introduce our action angle formulation of the osculating geodesic equations for generic Kerr orbits. 
We end this section by specialising to equatorial (i.e., spin aligned) orbits for the rest of this work.
Using these equations, along with a model for the gravitational self-force, allows us to calculate eccentric, self-forced, inspirals in Kerr spacetime for the first time.
However, these inspirals, henceforth referred to as ``osculating geodesic'' (OG) inspirals, are slow to compute, taking on the order of hours or days.

This motivates us to apply a near identity transformation, as developed in Paper I, which we summarize in Sec.~\ref{section:NITs}.
In Sec.~\ref{section:averaged_EoM} we explicitly derive the averaged equations of motion for the case of eccentric Kerr inspirals. 
Inspirals calculated with these equations of motion can be evaluated in less than a second, and are henceforth referred to as ``NIT'' inspirals.

For both OG and NIT inspirals, we require a model for the gravitational self-force. 
To this end, we review the calculation of the gravitational self-force in the radiation gauge in Sec.~\ref{section:GSF} before outlining a procedure used to interpolate this data in Sec.~\ref{section:Interpolation}. 
Using this interpolated self force model we describe our numerical implementation for calculating the various terms in the NIT equations of motion in Sec.~\ref{section:Implementation}. 

We then present the results of this implementation in Sec.~\ref{section:Results}.
We start with a consistency check with energy and angular momentum fluxes and an interpretation of the various terms in the NIT equations of motion in Sec.~\ref{section:Consistency_Checks}. 
We then compare OG and NIT inspirals to check that they agree at the appropriate order in the mass-ratio and assess the speed up the NIT provides.
We then explore some post-adiabatic effects of the first-order self force by comparing  NIT and adiabatic inspirals in Sec.~\ref{section:AdbVsPA}.
To round off our results, in Sec.~\ref{section:GaugeComparison} we make comparisons between inspiral trajectories around a Schwarzschild black hole calculated using two different first order self-force models: one calculated using outgoing radiation gauge self-force data and the other using Lorenz gauge  self-force data. 
These comparisons indicate that trajectories calculated using only first order self-force data are gauge dependent.
We end with some concluding remarks in Sec.~\ref{section:Discussion}.  

Throughout this paper we work in geometric units such that the gravitational constant and the speed of light are both equal to one (i.e., $G=c=1$).

\section{Forced motion near a rotating black hole} \label{section:EoM}

In this section we describe the motion of a non-spinning compact object of mass $\mu$ moving in the Kerr spacetime under the influence of some arbitrary force.
Later in this work, we will take this to be the self-force experienced by the secondary due to its interaction with its own metric perturbation.
We denote the mass of the primary by $M$ and parametrize its spin by $a = |J|/M$ where $J$ is the angular momentum of the black hole.
The Kerr metric can then be written in modified Boyer-Lindquist coordinates, $x^\alpha = \{t,r,z=\cos\theta,\phi\}$, as
\begin{equation}\label{eq:metric}
	\begin{split}
		ds^2 = & - \left(1 - \frac{2 M r}{\Sigma} \right) dt^2 + \frac{\Sigma}{\Delta} dr^2 + \frac{\Sigma}{1-z^2} dz^2 \\                 & + \frac{1-z^2}{\Sigma} (2a^2r (1 - z^2) + \Sigma \varpi^2) d\phi^2 - \frac{4 M a r (1-z^2)}{\Sigma}dt d\phi
	\end{split}
\end{equation}
where
\begin{subequations}\label{eq:DeltaSigmaOmega}
	\begin{gather}
		\Delta(r) := r^2 + a^2 - 2 M r,   \quad   \Sigma(r,z) := r^2 + a^2 z^2,   \quad   \varpi(r) := \sqrt{r^2 + a^2},    \tag{\theequation a-c}
	\end{gather}
\end{subequations}
If a force acts upon the secondary its motion can be described by the forced geodesic equation
 \begin{equation} \label{eq:forced_geodesic_eq}
	u^\beta\nabla_\beta u^\alpha = a^{\alpha}
\end{equation}
where $u^\alpha = dx^\alpha/d\tau$ is the secondary's four-velocity, $\nabla_\beta$ is the covariant derivative with respect to the Kerr metric, and $a^{\alpha}$ is the secondary's four-acceleration. 
We seek to recast Eq.~\eqref{eq:forced_geodesic_eq} into a form useful for applying the near-identity transformations.
Before considering the forced equation it is useful to first examine the geodesic limit.

\subsection{Geodesic motion and orbital parametrization} 
In the absence of any perturbing force, the secondary will follow a geodesic, i.e.,
\begin{equation} \label{eq:geodesic_eq}
	u^\beta\nabla_\beta u^\alpha = 0,
\end{equation} 
The symmetries of the Kerr spacetime allow for the identification of integrals of motion $\vec{\mathcal{P}} = \{ \En, \Lz, \K\}$ given by
\begin{subequations}\label{eq:ELK}
	\begin{gather}
		\En = - u_t,   \quad   \Lz = u_\phi,   \quad   \K = \K^{\alpha \beta} u_\alpha u_\beta,    \tag{\theequation a-c}
	\end{gather}
\end{subequations}
where $\K^{\alpha \beta}$ is the Killing tensor, $\En$ is the orbital energy per unit rest mass $\mu$, $\Lz$ is the z-component of the angular momentum divided by $\mu$ and $\K$ is the Carter constant divided by $\mu^2$  \cite{Carter1968}.
This definition of the Carter constant is related to another common definition of the Carter Constant, $\Q$, by 
\begin{equation}\label{eq:Carter_Constant}
	\Q = \K - (\Lz - a \En)^2.
\end{equation}
The geodesic equation can be written explicitly in terms of these integrals of motion \cite{Drasco2004}:
\begin{subequations}\label{eq:Geodesic_eqs}
	\begin{equation} \label{eq:Vr}
		\begin{split}
			\left( \frac{d r}{d\lambda} \right)^2 &= \left(\En \varpi^2 - a \Lz\right)^2 - \Delta\left(r^2+ \K \right) \\
			&= (1-\En^2)(r_1-r)(r-r_2)(r-r_3)(r-r_4) \coloneqq V_r,
		\end{split}
	\end{equation}
	\begin{equation}\label{eq:Vz}
		\begin{split}
			\left( \frac{d z}{d\lambda} \right)^2 &= \Q - z^2 \left(a^2 (1-\En^2)(1-z^2) + \Lz^2 + \Q \right)\\
			&= (z^2-z_-^2)\left(a^2 (1-\En^2)z^2 - z_+^2 \right) \coloneqq V_z,
		\end{split}
	\end{equation}
	
	\begin{equation}\label{eq:Geodesic_t}
		\begin{split}
			\frac{d t}{d\lambda} &= \frac{\varpi^2}{\Delta} \left(\En \varpi^2 - a \Lz \right) - a^2 \En(1-z^2)+ a \Lz \coloneqq s_t,
		\end{split}
	\end{equation}
	
	\begin{equation} \label{eq:Geodesic_phi}
		\begin{split}
			\frac{d \phi}{d\lambda} &= \frac{a}{\Delta} \left(\En \varpi^2 - a \Lz \right) + \frac{\Lz}{1-z^2} - a \En \coloneqq  s_{\phi}.
		\end{split}
	\end{equation}
\end{subequations}
where $r_1 > r_2 > r_3 > r_4$ are the roots of the radial potential $V_r$, $z_+ > z_-$ are the roots of the polar potential $V_z$, and $\lambda$ is Mino(-Carter) time that decouples the radial and polar equations \cite{Mino2003}. 
This time is related to the proper time of the particle, $\tau$, by
\begin{equation}\label{eq:Mino}
	d\tau = \Sigma d \lambda.
\end{equation}
The two largest roots of $V_r$ correspond to the apoapsis and periapsis of the orbit, respectively.
Explicit expressions for the other roots are derived in Ref.~\cite{Fujita2009a} and given for completeness in \ref{apdx:geodesic}.
Rather than parametrize an orbit by the set $\{\En, \Lz, \K\}$ it is useful to instead use more geometric, quasi-Keplerian constants $\vec{P} = \{p,e,x\}$.
Here $p$ is the semi-latus rectum, $e$ is the orbital eccentricity and $x$ measures the orbital inclination.
These are related to the radial and polar roots via
\begin{subequations}\label{eq:primary_roots}
	\begin{gather}
		r_1 = \frac{p M}{1-e},   \quad  r_2 = \frac{p M}{1+e},   \quad   z_- = \sqrt{1-x^2},    \tag{\theequation a-c}
	\end{gather}
\end{subequations}
The explicit relation between the integrals of motion $\{\En,\Lz,\K\}$ and $\{r_1,r_2,z_-\}$ can be found in, e.g., Appendix B of Ref.~\cite{Schmidt2002}.
We note that one advantage of using $x$ over other other common choices for inclination is that $x$ smoothly parametrizes orbits between prograde equatorial motion with $x=1$ to retrograde equatorial motion with $x=-1$ orbits.
Not all values of $\{p,e,x\}$ correspond to bound geodesics and we denote the value of $p$ at the last stable orbit by $p_{\text{LSO}}(a,e,x)$ \cite{Glampedakis2002,Stein2020}.

In order to later apply the near-identity transformations it will be useful to employ action-angle formulation to parametrize the orbital motion \cite{Schmidt2002,Fujita2009a,VandeMeent2020}.
In this description the orbital phases $\vec{q} = \{q_r,q_z\}$ are such that the geodesic equations can written in the form
\begin{subequations}\label{eq:GeodesicEqsPandq}
	\begin{gather}
		\dot{P_j} = 0\quad \text{and}  \quad \dot{q_i} =  \Upsilon_i (\vec{P}),   \tag{\theequation a-b}
	\end{gather}
\end{subequations}
where an overdot denotes derivative with respect to Mino time and the $\Upsilon_i$ are the Mino time fundamental frequencies \cite{Fujita2009a}.
Note that the right hand side of the $\dot{q}_i$ equation depends only on $\vec{P}$ and not also on the orbital phases.
Semi-analytic solutions to Eqs.~\eqref{eq:Geodesic_eqs} in terms of $\vec{q}$ can be found in Refs.~\cite{Fujita2009a,VandeMeent2020} and we present the key equations in \ref{apdx:geodesic}.

At this point we note that it is also common in the literature \cite{Drasco:2003ky, Nasipak:2019hxh} to express the radial and polar motion in terms of quasi-Keplerian angles, $\psi$ and $\chi$, via
\begin{subequations}\label{eq:Keplerian_Angles}
	\begin{gather}
		r(\psi) = \frac{p M}{ 1 + e \cos(\psi)} \quad \text{and}  \quad z(\chi) = z_- \cos (\chi).    \tag{\theequation a-b}
	\end{gather}
\end{subequations}
With this parametrization the evolution equations for $\psi$ and $\chi$ depend on both $\vec{P}$ and $\vec{q}$.
This makes them inconvenient for deriving averaging transformations, though it is not an insurmountable challenge \cite{Pound2021}. 
For this reason we prefer the action-angle formulation.

\subsection{Osculating Geodesics} \label{section:Full_EoM}

We now wish to describe the forced motion of a body obeying Eq.~\eqref{eq:forced_geodesic_eq}.
To do so, we make use of the method of osculating orbital elements, or osculating geodesics when applied to the relativistic context \cite{Pound2008,Gair2011}. 
We first identify a set of orbital elements that uniquely identify a given geodesic, such as the integrals of motion $\vec{P}$ along with the initial values of the orbital phases of the geodesic orbit $\vec{q}_0$ and designate them as a set of ``orbital elements" $\vec{I} = \{\vec{P},\vec{q}_0\}$.
For accelerated orbits, these orbital elements are promoted from constants to functions of Mino time. Note that now the orbital elements $\vec{q}_0(\lambda)$ are different quantities from the values of $\vec{q}$ evaluated at $\lambda = 0$, i.e., $\vec{q}(0)$.
We assume that the worldline and four-velocity of the secondary at each instant can be described as the worldline and four-velocity of a test body on a tangent geodesic, i.e.,  $x^\alpha (\lambda) = x_G^\alpha (I^A(\lambda),\lambda )$ and $u^\alpha(\lambda) = u_G^\alpha(I^A(\lambda),\lambda)$.
From these assumptions, one can derive the osculating geodesic equations \cite{Pound2008,Gair2011}:
\begin{subequations}\label{eq:osculating_equations}
	\begin{gather}
		\frac{\partial x^{\alpha}_G}{\partial I^A} \dot{I}^A = 0  
		\quad \text{and} \quad  
		 \frac{\partial u^G_{\alpha}}{\partial I^A} \dot{I}^A = \Sigma a_{\alpha}.   
		 \tag{\theequation a-b}
	\end{gather}
\end{subequations}
From these, evolution equations for each of the orbital elements can be calculated, which we have done in \ref{EvolutionOfCoM} and \ref{EvolutionOfPhases}.
This was first done for generic Kerr orbits in Ref.~\cite{Gair2011} which lays out four different formulations of the equations. 
Three of these formulations use the quasi-Keplerian angles $\psi$ and $\chi$ as the orbital phases, and two of these were numerically implemented and shown to agree with each other. 
We make use of the final formulation where one instead uses the geodesic actions angels $q_r$ and $q_z$ as the orbital phases.

We first find the evolution equations for the integrals of motion $\vec{P}$. 
We do so by finding evolution equations for $\vec{\mathcal{P}}$ and relating these to evolution equations for the roots $r_1$, $r_2$ and $z_-$. 
We derive these relations in \ref{EvolutionOfCoM}. 
From here, one can invert Eqs.~\eqref{eq:primary_roots} to find
\begin{subequations}\label{eq:OrbitalElementEvolutionEqs}
	\begin{equation}
		\frac{ d p}{d \lambda} = \frac{2}{M (r_1 + r_2)^2} \left( r_2^2\frac{d r_1}{d \lambda} + r_1^2 \frac{d r_2}{d \lambda} \right) \coloneqq F_p ,
	\end{equation}
	\begin{equation}
		\frac{ d e}{d \lambda} = \frac{2}{(r_1 + r_2)^2} \left( r_2\frac{d r_1}{d \lambda} - r_1 \frac{d r_2}{d \lambda} \right) \coloneqq F_e,
	\end{equation}
	\begin{equation}
		\frac{ d x}{d \lambda} = - \frac{z_-}{x}\frac{d z_-}{d \lambda} \coloneqq F_x.
	\end{equation}
\end{subequations}

The orbital phases $\vec{q}$ still evolve with their respective Mino-time frequencies, but now pick up a correction due to the evolution of the initial values, i.e.,
\begin{equation}
	\frac{d q_{i}}{d \lambda} =  \Upsilon_i + \frac{d q_{i,0}}{d \lambda}.
\end{equation}
To find the evolution equations for the initial values for the orbital phases, we can re-arrange the first osculating condition (\ref{eq:osculating_equations}a) and exploit the fact that the evolution of $r$ is independent of $q_z$, and the evolution $z$ is independent of $q_r$, to get
\begin{equation}\label{eq:phase_eq_1}
	\frac{d q_{i,0}}{d \lambda} =  - \frac{1}{\partial x_G^i/ \partial q_i} \left(\frac{\partial x_G^i}{\partial P_j} \frac{d P_j}{d \lambda}\right) \coloneqq f_i^{(1)},
\end{equation}
where $x_G^i$ is the geodesic expression for $r$ or $z$ given by Eqs.~\eqref{eq:r_analytic} and \eqref{eq:z_analytic} respectively.
Unfortunately, this expression is difficult to evaluate numerically at the orbital turning points where both $\partial x_G^i/ \partial q_i$ and the term in the parentheses go to zero.
However, one can derive an alternative expression for this quantity that is regular at the turning points \cite{Gair2011}: 
\begin{equation}\label{eq:phase_eq_2}
	\frac{d q_{i,0}}{d \lambda} =   \frac{ 2\Upsilon_i}{\partial V_i / \partial x_G^i} \left( \Sigma^2 a^i - \left(\frac{\partial \dot{x}_G^i}{\partial P_j} \dot{P_j} \right)\right) \coloneqq f_i^{(1)}.
\end{equation}
See \ref{EvolutionOfPhases} for details on the derivation. 
This expression instead has a singularity whenever $\partial V_i / \partial x_G^i = 0$. 
Thus, for our numerical implementation, we use Eq.~\eqref{eq:phase_eq_1} for the majority of the orbital cycle and switch to Eq.~\eqref{eq:phase_eq_2} in the vicinity of turning points.

Finally, we also require evolution equations for ``extrinsic quantities" that don't show up on the right hand side of the equations of motion, but are necessary to compute the trajectory and the waveform. 
These are the time and azimuthal coordinates of the secondary which, as a set, we denote by $\vec{S} = \{t,\phi\}$.
Their evolution is given by the geodesic equations for $t$ and $\phi$, i.e., equations (\ref{eq:Geodesic_eqs}c) and (\ref{eq:Geodesic_eqs}d). 
Putting it all together, the equations of motion take the form:
\begin{subequations}\label{eq:Generic_EMRI_EoM}
	\begin{align}
		\begin{split}
			\dot{P_j} &= F_j (\vec{P}, \vec{q}) ,
		\end{split}\\
		\begin{split}
			\dot{q_i} &=  \Upsilon_i (\vec{P}) +  f_i (\vec{P}, \vec{q}),
		\end{split}\\
		\begin{split}
			\dot{S_k} &= s_k(\vec{P}, \vec{q}).
		\end{split}
	\end{align}
\end{subequations}
These equations of motion are valid for generic inspirals under the influence of an unspecified perturbing force.
We find that the action angle implementation produces inspiral trajectories that are identical to inspirals calculated using the ``null tetrad" formulation described in Ref.~\cite{Gair2011}. 
We have implemented both the action angle and null tetrad osculating element equations into a Mathematica package that will be publicly available on the Black Hole Perturbation Toolkit \cite{BHPToolkit}.
We find numerically that the null tetrad formulation is more computationally efficient as it does not have any singular equations that necessitate switching between different formulations twice during each orbital period.
As such, for direct comparisons between OG and NIT inspirals and waveforms we make use of the null tetrad formulation, but use the action angle formulation as the starting point for our averaging procedure.

\subsection{Specialising to equatorial motion}

For the rest of this work, we specialize to the case of eccentric inspirals in the equatorial plane (i.e., spin aligned) under the influence of the first-order ratio gravitational self force (GSF). 
This corresponds to setting $x = \pm 1$ for prograde and retrograde orbits, respectively. 
Due to symmetry, motion in the equatorial plane will stay in the equatorial plane, and thus $\dot{x} = 0$.
As such, we only need to track the evolution of $\vec{P} = \{p,e\}$. 
Similarly, the equations of motion no longer depend on the polar phase $q_z$, and so we only need to evolve the radial phase $\vec{q} = \{q_r\}$.
The gravitational self force scales with the small mass ratio $\mr = \mu/M$, meaning that the secondary's four-acceleration can be expressed as $a^\alpha =  \mr a^\alpha_{\text{GSF}}$. 
Factoring out this scaling, the equations of motion for equatorial inspirals become
\begin{subequations}\label{eq:Equatorial_EoM}
	\begin{align}
		\begin{split}
			\frac{d p}{d \lambda} &= \mr F_p (a, p, e, q_r),
		\end{split}\\
		\begin{split}
			\frac{d e}{d \lambda} &= \mr F_e (a, p, e, q_r),
		\end{split}\\
		\begin{split}
			\frac{d q_r}{d \lambda} &=  \Upsilon_r(a, p,e) + \mr f_r (a, p, e, q_r),
		\end{split}\\
		\begin{split}
			\frac{d t}{d \lambda} &= s_t(a, p, e, q_r), 
		\end{split}\\
		\begin{split}
			\frac{d \phi}{d \lambda} &= s_\phi(a, p, e, q_r).
		\end{split}
	\end{align}
\end{subequations}

\section{Near-Identity Transformations} 
Near identity (averaging) transformations (NITs) are well known technique in applied mathematics and celestial mechanics \cite{Kevorkian1987}.
This technique involves making small transformations to the equations of motion, such that the short timescale physics is averaged out, while retaining information about the long term evolution of a system. 
This is well suited to modelling EMRIs, where we do not require perfect knowledge of the trajectory on the orbital timescale, so long as we can accurately track its evolution on the radiation reaction timescale. 
We note that there can be a relationship between gauge transformations and NITs, as explored in Ref.~\cite{Fujita:2016igj}, the NITs we perform in this work are distinct from the choice of gauge \cite{Pound:2015fma}.
In Paper I \cite{NITs}, these transformations were derived for EMRIs in general and then applied to the OG equations of motion around a Schwarzschild black hole. 
We briefly review the work of Paper I for a general EMRI before applying these transformations to the specific case of eccentric self-forced inspirals in Kerr.

\subsection{Near identity averaging transformations for generic EMRI systems} \label{section:NITs}

The NIT variables, $\nit{P}_j$, $\nit{q}_i$ and $\nit{S}_k$, are related to the OG variables  $P_j$, $q_i$ and $S_k$ via
\begin{subequations}\label{eq:transformation}
	\begin{equation}\label{eq:transformation1}
		\nit{P}_j = P_j + \sp Y_j^{(1)}(\vec{P},\vec{q}) + \sp^2 Y_j^{(2)}(\vec{P},\vec{q}) + \HOT{3},
	\end{equation}
	\begin{equation}
		\nit{q}_i = q_i + \sp X_i^{(1)}(\vec{P},\vec{q}) +\sp^2 X_i^{(2)}(\vec{P},\vec{q}) + \HOT{3},
	\end{equation}
	\begin{equation}
		\nit{S}_k = S_k + Z_k^{(0)}(\vec{P},\vec{q}) +\sp Z_i^{(1)}(\vec{P},\vec{q}) + \HOT{2}.
	\end{equation}
\end{subequations}
Here, the transformation functions $Y_j^{(n)}$, $X_i^{(n)}$, and $Z_k^{(n)}$ are required to be smooth, periodic functions of the orbital phases $\vec{q}$. 
At leading order, Eqs.~\eqref{eq:transformation} are identity transformations for $P_k$ and $q_i$ but not for $S_k$ due to the presence of a zeroth order transformation term $Z_k^{(0)}$. 
The inverse transformations can be found for $P_k$ and $q_i$ by requiring that their composition with the transformations in Eqs.\eqref{eq:transformation} must give the identity transformation. Expanding order by order in $\epsilon$, this gives us
\begin{subequations}\label{eq:inverse_trasformaiton}
	\begin{equation}
		\begin{split}
			P_j &= \nit{P_j} - \epsilon Y_j^{(1)}(\vec{\nit{P}},\vec{\nit{q}}) \\ & - \epsilon^2 \left(  Y_j^{(2)}(\vec{\nit{P}},\vec{\nit{q}}) - \PD{Y_j^{(1)}(\vec{\nit{P}},\vec{\nit{q}})}{\nit{P_k}} Y_k^{(1)}(\vec{\nit{P}},\vec{\nit{q}}) - \PD{Y_j^{(1)}(\vec{\nit{P}},\vec{\nit{q}})}{\nit{q_k}} X_k^{(1)}(\vec{\nit{P}},\vec{\nit{q}})  \right) + \mathcal{O}(\epsilon^3),
		\end{split}
	\end{equation}
	\begin{equation}
		\begin{split}
			q_i &= \nit{q_i} - \epsilon X_i^{(1)}(\vec{\nit{P}},\vec{\nit{q}})  \\ &- \epsilon^2 \left(  X_i^{(2)}(\vec{\nit{P}},\vec{\nit{q}}) - \PD{X_i^{(1)}(\vec{\nit{P}},\vec{\nit{q}})}{\nit{P_j}} Y_j^{(1)}(\vec{\nit{P}},\vec{\nit{q}}) - \PD{X_i^{(1)}(\vec{\nit{P}},\vec{\nit{q}})}{\nit{q_k}} X_k^{(1)}(\vec{\nit{P}},\vec{\nit{q}})  \right) + \mathcal{O}(\epsilon^3).
		\end{split}
	\end{equation}
\end{subequations}

To proceed it is useful to decompose various functions into Fourier series where we use the convention:
\begin{equation} \label{eq:Fourier}
	A(\vec{P},\vec{q}) = \sum_{\vec{\kappa} \in \mathbb{Z}^N} A_{\vec{\kappa}}(\vec{P}) e^{i \vec{\kappa} \cdot \vec{q}},
\end{equation}
where $N$ is the number of orbital phases.
Based on this, we can split the function into an averaged piece
\begin{equation}\label{eq:average}
	\avg{A} (\vec{P}) = A_{\vec{0}}(\vec{P}) = \frac{1}{(2\pi)^N} \idotsint_{\vec{q}} A(\vec{P},\vec{q}) d q_1 \dots d q_N,
\end{equation}
and an oscillating piece
\begin{equation}\label{eq:oscillating}
	\osc{A}(\vec{P},\vec{q}) = A(\vec{P},\vec{q}) - \avg{A}(\vec{P})  = \sum_{\vec{\kappa} \neq \vec{0}} A_{\vec{\kappa}}(\vec{P}) e^{i \vec{\kappa} \cdot \vec{q}}. 
\end{equation}
Using the above transformations along with the equations of motion, and working order by order in $\mr$, we can chose values for the transformation functions $Y_j^{(1)}, Y_j^{(2)}, X_i^{(1)}, X_i^{(2)}, Z_k^{(0)}$ and $Z_k^{(1)}$ such that the resulting equations of motion for $\nit{P_j}, \nit{q_i}$ and $\nit{S_k}$ take the following form
\begin{subequations}\label{eq:transformed_EoM}
	\begin{equation}
		\dot{\nit{P}}_j = 0 + \epsilon \nit{F}_j^{(1)}(\vec{\nit{P}}) + \epsilon^2 \nit{F}_j^{(2)}(\vec{\nit{P}}) +  \mathcal{O}(\epsilon^3),
	\end{equation}
	\begin{equation}
		\dot{\nit{q}}_i = \Upsilon_i(\vec{\nit{P}}) +\epsilon \nit{f}_i^{(1)}(\vec{\nit{P}}) +\epsilon^2 \nit{f}_i^{(2)}(\vec{\nit{P}})+  \mathcal{O}(\epsilon^3),
	\end{equation}
	\begin{equation}
		\dot{\nit{S}}_k = \nit{s}_k^{(0)}(\vec{\nit{P}}) + \epsilon \nit{s}_k^{(1)}(\vec{\nit{P}}) +  \mathcal{O}(\epsilon^2).
	\end{equation}
\end{subequations}
Crucially, these equations of motion are now independent of the orbital phases $\vec{q}$. Deriving the relationship between the transformed forcing functions ($\nit{F}_j^{(1\backslash 2)},\nit{f}_i^{(1\backslash2)} $ and $\nit{s}_k^{(0\backslash1)}$) to the original forcing functions is quite an involved process with several freedoms and choices, each with their own merits and drawbacks. This is discussed at length in \cite{NITs}, so for brevity we will summarize the results and the particular choices we have made in this work. 

The transformed forcing functions are related to the original functions by 
\begin{subequations}
	\begin{equation}
		\nit{F}_j^{(1)} = \left<F_{j}^{(1)}\right>,
	\end{equation}
	\begin{equation}
		\nit{f}_i^{(1)}= \left<f_{i}^{(1)}\right>,
	\end{equation}
	\begin{equation}
		\nit{s}_k^{(0)} =  \left<s_{k}^{(0)}\right>,
	\end{equation}
	\begin{equation}
		\nit{F}_j^{(2)} = \left<F_j^{(2)} \right> + \left<\frac{\partial \osc{Y}_j^{(1)}}{\partial \nit{q}_i} \osc{f}_i^{(1)} \right> + \left<\frac{\partial \osc{Y}_j^{(1)}}{\partial \nit{P}_k} \osc{F}_k^{(1)} \right>,
	\end{equation}
	\begin{equation}
		\nit{f}_i^{(2)} = 0, \text{ and} 
	\end{equation}
	\begin{equation}
		\nit{s}_k^{(1)} = \left<s_k^{(1)} \right> - \left<\frac{\partial \osc{s}_k^{(0)}}{\partial \nit{P}_j} \osc{Y}_j^{(1)} \right> - \left<\frac{\partial \osc{s}_k^{(0)}}{\partial \nit{q}_i} \osc{X}_i^{(1)} \right>.
	\end{equation}
\end{subequations}
In deriving these equations of motion, we have constrained the oscillating pieces of the NIT transformation functions to be
\begin{equation}\label{eq:NIT_Y}
	\osc{Y}_j^{(1)} \equiv \sum_{\vec{\kappa} \neq \vec{0}} \frac{i}{\vec{\kappa} \cdot \vec{\Upsilon}} F_{j,\vec{\kappa}}^{(1)} e^{i \vec{\kappa} \cdot \vec{q}},
\end{equation}

\begin{equation}\label{eq:NIT_X}
	\osc{X}_i^{(1)} \equiv \sum_{\vec{\kappa} \neq \vec{0}}\left( \frac{i}{\vec{\kappa} \cdot \vec{\Upsilon}} f_{i,\vec{\kappa}}^{(1)} + \frac{1}{(\vec{\kappa} \cdot \vec{\Upsilon})^2} \frac{\partial \Upsilon_i}{\partial P_j}F_{j,\vec{\kappa}}^{(1)} \right) e^{i \vec{\kappa} \cdot \vec{q}},
\end{equation}
and $\osc{Z}_k^{(0)}$ is found by solving
\begin{equation}\label{eq:Z_definition}
	\osc{s}_k^{(0)} + \frac{\partial \osc{Z}_k^{(0)}}{\partial \nit{q}_i} \Upsilon_i = 0.
\end{equation}
This equation is satisfied by the oscillating pieces for the analytic solutions for the geodesic motion of $t$ and $\phi$ in Eqs.~\eqref{eq:Extrinsic_Analytic_Solutions}, i.e.,
\begin{equation}\label{eq:Z_solution}
	\osc{Z}_k^{(0)} = - \osc{S}_{k,r} (q_r) - \osc{S}_{k,z} (q_z).
\end{equation}
We have chosen the averaged pieces such that $\bavg{Y_j^{(1)}} = \bavg{Y_j^{(2)}} = \bavg{X_i^{(2)}} =\bavg{Z_k^{(0)}} =\bavg{Z_k^{(1)}} = 0$ but have used $\bavg{X_i^{(1)}}$ to cancel out the contributions of $\nit{f}_i^{(2)}$. 
In order to generate waveforms, one only needs to know the transformations in Eq.~\eqref{eq:transformation} to zeroth order in the mass ratio, i.e.,
\begin{subequations}
	\begin{equation}
		P_j = \nit{P_j} + \mathcal{O}(\epsilon),
	\end{equation}
	\begin{equation}
		q_i = \nit{q_i} + \mathcal{O}(\epsilon),
	\end{equation}
	\begin{equation}
		S_k = \nit{S_k} -Z_k^{(0)}(\vec{\nit{P}},\vec{\nit{q}} ) +  \mathcal{O}(\epsilon).
	\end{equation}
\end{subequations}
Furthermore, to be able to directly compare between OG and NIT inspirals, we will need to match their initial conditions to sufficient accuracy. To maintain an overall phase difference of $\mathcal{O}(\epsilon)$ in the course of an inspiral, this requires known the transformation of the $P_j$'s \eqref{eq:transformation1} to linear order in $\epsilon$, while it is sufficient to know the rest of Eq.~\eqref{eq:transformation} to zeroth order. In particular, we will never need an explicit expression for $\bavg{X_i^{(1)}}$, which would require solving a PDE to obtain.

\subsection{Averaged Equations of Motion for Eccentric Equatorial Kerr Inspirals} \label{section:averaged_EoM}
We now apply the near identity averaging transformation procedure to the equations of motion for equatorial Kerr inspirals to obtain:
\begin{subequations}\label{eq:NIT_EoM}
	\begin{align}
		\begin{split}
			\frac{d \nit{p}}{d \lambda} &= \mr \nit{F}_p^{(1)} (a, \nit{p}, \nit{e}) + \mr^2 \nit{F}_p^{(2)} (a, \nit{p}, \nit{e}),
		\end{split}\\
		\begin{split}
			\frac{d \nit{e}}{d \lambda} &= \mr \nit{F}_e^{(1)} (a, \nit{p}, \nit{e}) + \mr^2 \nit{F}_e^{(2)} (a, \nit{p}, \nit{e}),
		\end{split}\\
		\begin{split}
			\frac{d \nit{q_r}}{d \lambda} &=  \Upsilon_r(a, \nit{p},\nit{e}) + \mr \nit{f}_r^{(1)} (a, \nit{p}, \nit{e}),
		\end{split}\\
		\begin{split}
			\frac{d \nit{t}}{d \lambda} &= \nit{s}_t^{(0)}(a, \nit{p}, \nit{e}) + \mr \nit{s}_t^{(1)}(a, \nit{p}, \nit{e}), 
		\end{split}\\
		\begin{split}
			\frac{d \nit{\phi}}{d \lambda} &= \nit{s}_{\phi}^{(0)}(a, \nit{p}, \nit{e}) + \mr \nit{s}_{\phi}^{(1)}(a, \nit{p}, \nit{e}).
		\end{split}
	\end{align}
\end{subequations}
The leading order terms in each equation of motion are simply the original function averaged over a single geodesic orbit, i.e.,
	\begin{equation} \label{eq:NIT_Relationship1}
		\nit{F}_p^{(1)} = \avg{F_{p}}, \quad \nit{F}_e^{(1)} = \avg{F_{e}}, \quad \nit{f}_r^{(1)}= \avg{f_{r}},
	\end{equation}
	\begin{equation}\label{eq:NIT_Relationship2}
		\nit{s}_t^{(0)} =  \avg{s_{t}} = \Upsilon_t,\quad \nit{s}_{\phi}^{(0)} =  \avg{s_{\phi}} = \Upsilon_{\phi} , 
	\end{equation}
where $\Upsilon_t$ and $\Upsilon_\phi$ are the Mino-time $t$ and $\phi$ fundamental frequencies. 
The remaining terms are more complicated and require Fourier decomposing the original functions and their derivatives with respect to to the orbital elements $(p,e)$. To express the result, we define the operator
\begin{equation} \label{eq:N_Operator}
	\mathcal{N}(A) =  \sum_{\kappa \neq 0} \frac{-1}{\Upsilon_r} 
	\biggl[ A_{\kappa} f_{r,-\kappa} - 
	\frac{i}{\kappa} \left( \PD{A_{\kappa} }{\nit{p}} F_{p,-\kappa} +
	\PD{A_{\kappa} }{\nit{e} } F_{e,-\kappa} -
	\frac{A_{\kappa}}{\Upsilon_r} \left( \PD{\Upsilon_r}{\nit{p}} F_{p,-\kappa} +
	\PD{\Upsilon_r}{\nit{e}} F_{e,-\kappa} \right) \right) \biggr]
\end{equation}
With this in hand, the remaining terms in the equations of motion are found to be
	\begin{equation}\label{eq:NIT_Relationship3}
		\nit{F}_p^{(2)} =  \mathcal{N}(F_p), \quad \nit{F}_e^{(2)} =  \mathcal{N}(F_e), \quad \nit{s}_t^{(1)} =  \mathcal{N}(s_t), \quad \nit{s}_\phi^{(1)} =  \mathcal{N}(s_\phi).
	\end{equation}

 \section{Gravitational self-force for eccentric Kerr inspirals}

In order to drive the inspiral we need to rapidly evaluate the gravitational self-force (GSF) given any values of $(p,e,q_r)$.
Codes that compute the GSF generally take minutes to hours to compute the force along a geodesic for a given $(p,e)$ value and so it is not practical to directly couple the equations of motion to a GSF code.
Instead, it is common practice to calculate the GSF on a discrete set of points across the parameter space and then build an interpolation or fitted model that smoothly connects the GSF data.
The following subsections describe our approach.
 
 \subsection{Gravitational self-force} \label{section:GSF}
 
 The gravitational self-force approach is reviewed extensively elsewhere \cite{Barack:2018yvs, Poisson:2011nh} and so we just give a brief overview of the calculations that we employ.
 The GSF approach starts by expanding the metric of the binary around the metric of the primary, i.e., $g_{\mu\nu} = \bar{g}_{\mu\nu} + \epsilon h^{(1)}_{\mu\nu} +\epsilon^2 h^{(2)}_{\mu\nu} + \dots$ where $\bar{g}_{\mu\nu}$ is the Kerr metric and the $h^{(n)}$ are the \mbox{$n$-th} order perturbations to the spacetime due to the presence of the secondary. 
 The interaction between these metric perturbations and the motion of the secondary can be derived through a matched asymptotic expansion analysis \cite{Poisson:2011nh}.
 In this work we use only first-order (in the mass ratio) results, as second order results are still emerging \cite{Warburton2021}. 
 The first order metric perturbation generated by a compact object can be found by solving the linearized Einstein Field Equations with a point particle source moving on a geodesic of $\bar{g}_{\mu\nu}$. The matched asymptotic expansion analysis identifies a regular part of this (divergent) metric perturbation that is responsible for the backreaction on the compact object \cite{Mino:1996nk,Quinn:1996am, Detweiler:2002mi,Poisson:2011nh}.
 The result is a (self-)force that appears on the right-hand side of Eq.~\eqref{eq:forced_geodesic_eq} computed from derivatives of the regular metric perturbation \cite{Poisson:2011nh}.
 
 Solving the perturbation equations requires picking a gauge, and the resulting self-force is gauge dependent \cite{Barack:2001ph}.
 The self-force was first computed in the Lorenz gauge where the procedure for obtaining the regular part was best understood.
 Numerical calculations of the Lorenz gauge self-force have been made in both the frequency- \cite{Akcay:2010dx,Akcay2013,Osburn2014} and time-domains \cite{Barack2007,Barack2010b,Dolan2013}
 All these results have been for motion in Schwarzschild spacetime, with one exception in Kerr~\cite{Isoyama:2014mja}.
 
 Calculating perturbations of the Kerr spacetime is hampered by the lack of separability of the linearized Einstein Field Equations on this background. This difficulty can be circumvented by using the Teukolsky formalism for describing perturbations to the Weyl scalars \cite{Teukolsky:1973ha}, which is fully separable in the frequency domain. From the Weyl scalars, the metric perturbation can be reconstructed in a radiation gauge \cite{Chrzanowski:1975wv,Kegeles:1979an, Wald:1978vm}.
 There has also been recent progress understanding how to reconstruct the metric in the Lorenz gauge \cite{Dolan:2021ijg}. 
 Regularization of the metric perturbation in radiation gauges is more subtle \cite{Pound:2013faa}, but self-force calculations in the radiation gauge are now routine \cite{Merlin:2014qda,vandeMeent:2015lxa,VandeMeent2016,VanDeMeent2018}. 
 
 In this work we primarily use the self-force computed using the code of Ref.~\cite{vandeMeent:2015lxa,VandeMeent2016,VanDeMeent2018}.
 This code uses the Mano-Suzuki-Takasugi methods \cite{Mano:1996vt,Mano:1996gn,Fujita:2004rb,Fujita:2009us} to compute the perturbations to the Weyl scalars in the frequency domain.
 The metric is then reconstructed into an outgoing radiation gauge (including mass and angular momentum perturbations \cite{Merlin:2016boc,vandeMeent:2017oet} and gauge completion contributions \cite{gaugecompletion}).
 The metric perturbation is then projected onto a basis of spherical harmonics before regularization is carried out using the mode-sum approach \cite{Barack:1999wf,Pound:2013faa}.
 Depending on the eccentricity of the orbit the code must compute the metric perturbation by summing over thousands to tens of thousands of Fourier harmonic modes.
 With the current Mathematica implementation, the self-force for an, e.g., $a=0.9M, p=3.375, e=0.5$ orbit takes approximately 90 CPU hours to compute.

 The self-force can be split into dissipative (time anti-symmetric) and conservative (time-symmetric) contributions \cite{Barack:2009ux}.

 The dissipative pieces causes the orbit to shrink until the secondary plunges into the primary.
 It also generally causes the orbit to circularize, with the exception being just before the transition to plunge where the orbit gains eccentricity \cite{Cutler:1994pb,Glampedakis2002,Fujita:2020zxe,Isoyama:2021jjd}.
 To produce adiabatic waveforms, we only require knowledge of the orbit averaged dissipative pieces of the first-order self-force.
 These can be related, via balance laws, to the fluxes of GWs to infinity and down the event horizon.
 Since calculating fluxes avoids regularization of the metric perturbation, adiabatic inspirals are typically calculated via flux balance laws \cite{Hughes2000,Hughes2001,Glampedakis2002,Hughes2005,Fujita:2020zxe,Isoyama:2021jjd, Hughes2021}.
 The conservative pieces have more subtle effects on the inspiral, such as altering the rate of periapsis advance and the location of the innermost stable circular orbit \cite{Barack2009,Barack2010a,Barack2011,Warburton2011,VanDeMeent2017, Vines:2015efa, Fujita:2016igj}.

 To compute post-adiabatic inspirals requires knowledge of both the dissipative and conservatives pieces of the first-order self-force and the orbit average piece of the second-order self-force \cite{Hinderer2008}.
 There are as yet no calculations of the latter so we will make do with just the first-order self-force information in this work.
 This will allow us to explore some of the effects of the conservative self-force on equatorial Kerr inspirals for the first time. 
 
 \subsection{Interpolation method} \label{section:Interpolation}
 
 Driving inspirals requires a model for the self-force that can be rapidly evaluated at each instant during the inspiral.
 To achieve this we tile the parameter space with GSF data which we can then interpolate. 
 While there have been many different approaches for doing this, we has been implemented for eccentric Schwarzschild inspirals \cite{Warburton2012, Osburn2016}.
 In these works the data was interpolated using standard cubic spline methods, which required computing the self-force at tens of thousands of points in the parameter space.
 While this might not pose much of a problem for the 2D parameter space of eccentric, Schwarzschild inspirals, these approaches would not scale well to the 4D parameter space for generic Kerr inspirals.
 Motivated by this, as well as the computational expense of the eccentric Kerr self-force code, we build an interpolation model based on Chebyshev polynomials that is accurate to percent level across a 2D slice of the EMRI parameter space using only a few hundred points.
 
 We start by fixing the value of the spin parameter of the primary, which we choose to be $a = 0.9M$ for Kerr inspirals or $a = 0$ for Schwarzschild inspirals and set the inclination $x$ to be either $1$ or $-1$ for prograde orbits or retrograde orbits respectively. 
 This reduces our parameter space to two parameters; the semilatus rectum $p$ and the eccentricity $e$. 
 We then define a parameter $y$ using the $p$ and the position of the last stable orbit $p_{\text{LSO}}$. 
 For Kerr orbits, we chose $y$ to be 
 \begin{equation}\label{eq:y_Kerr}
 	y_{\text{Kerr}} = \sqrt{\frac{p_{\text{LSO}}(a,e,x)}{p}}.
 \end{equation}
With this parametrization we found that the accuracy of the Chebyshev interpolation is limited by the appearance of cusps at the LSO in the data. To ameliorate their impact we instead used a parameter $y$ given by
 \begin{equation}\label{eq:y_Schwarzschild}
 	y_{\text{Schwarz}} = 1 - \left( 1- \frac{p_{\text{LSO}}(0,e,x)}{p} \right)^{1/3}
 \end{equation}
for later runs in Schwarzschild spacetime. 
In either case, tiling the parameter space in $y$ instead of $p$ will concentrate more points near the separatrix where the self force varies the most.
 
We let $y$ range from $y_{\text{min}} = 0$ ($0.01$ for Schwarzschild) to $y_{\text{max}} = 1$ and $e$ range from $e_{\text{min}} = 0$ to $e_{\text{max}} = 0.5$ for Kerr and $e_{\text{min}} = 0$ to $e_{\text{max}} = 0.3$ for Schwarzschild.
We define parameters $u$ and $v$ which cover this parameter space as they range from $(-1,1)$
\begin{subequations}
	\begin{gather}
		u \coloneqq \frac{y-(y_{\text{min}}+y_{\text{max}})/2}{(y_{\text{min}}-y_{\text{max}})/2}
		\quad \text{and} \quad  
		v \coloneqq \frac{e-(e_{\text{min}}+e_{\text{max}})/2}{(e_{\text{min}}-e_{\text{max}})/2},   
		\tag{\theequation a-b}
	\end{gather}
\end{subequations}
 This parametrization is convenient when using Chebyshev polynomials of the first kind, where the order $n$ polynomial is defined by $T_n (\cos \varphi) \coloneqq \cos (n \varphi)$.
 The Chebyshev nodes are the roots these polynomials, and the location of the $k$th root of $n$th polynomial is given by
 \begin{equation}
 N_k = \cos{\left( \frac{2 k -1}{2n} \pi \right)}
  \end{equation}
We then calculate the GSF on a $15 \times 7$ grid of Chebyshev nodes, with the $u$ values given by the roots of the 15th order polynomial and the $v$ values given by the roots of the 7th order polynomial.
At each point on our grid, we Fourier decompose each component of the force with respect to the radial action angle $q_r$.
 We then multiply the data for each Fourier coefficient by a factor of $(1-y)/(1-e^2)$, as we find that this smooths the behaviour of the force near the separatrix and improves the accuracy of our interpolation.
 Next, we use Chebyshev polynomials to interpolate each Fourier coefficient across the $(u,v)$ grid. 
 We then resum the modes to reconstruct our interpolated gravitational self force model:
 
 	\begin{equation}
 		a_\alpha = \frac{1-e^2}{1-y} \sum_{\kappa = 0}^{15} A^\kappa_\alpha(y,e) \cos (\kappa q_r) + B^\kappa_\alpha(y,e) \sin (\kappa q_r),
 	\end{equation}
	where
 	\begin{equation}
 		A^\kappa_\alpha(y,e) = \sum_{i = 0}^{14}  \sum_{j = 0}^{6} A^{\kappa i j}_{\alpha} T_i \left(u\right) T_j\left(v\right)
 		\quad \text{and} \quad
 		B^\kappa_\alpha(y,e) = \sum_{i = 0}^{14}  \sum_{j = 0}^{6} B^{\kappa i j}_{\alpha} T_i \left(u\right) T_j\left(v\right)
 	\end{equation}
	Using this procedure forces each component to become singular at the last stable orbit.
	While the GSF changes rapidly as one approaches the last stable orbit, we do not expect the components of the self force to diverge at the LSO. 
	Understanding the analytic structure of the self-force in this region would likely improve future interpolation models.

 We note that the GSF should satisfy the orthogonality condition with the geodesic four-velocity, i.e., $a_{\alpha} u^{\alpha} = 0$.
 Interpolation will bring with it a certain amount of error which can cause this condition to be violated. 
 We find empirically that we can  reduce this interpolation error by projecting the force so that this condition is always satisfied, i.e.,
 \begin{equation}
 	a_\alpha^\perp = a_\alpha + a_\beta u^\beta u_\alpha.
 \end{equation}
 
 \begin{figure} 
 	\centering
 	\begin{subfigure}[b]{0.496\textwidth}
 		\centering
 		\includegraphics[width=\textwidth]{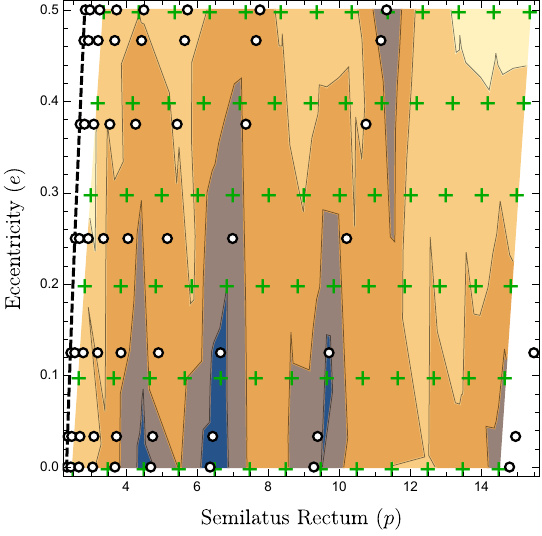}
 		\caption{$a_t$}
 		\label{fig:FtInterpolationError}
 	\end{subfigure}
 	\hfill
 	\begin{subfigure}[b]{0.496\textwidth}
 		\centering
 		\includegraphics[width=\textwidth]{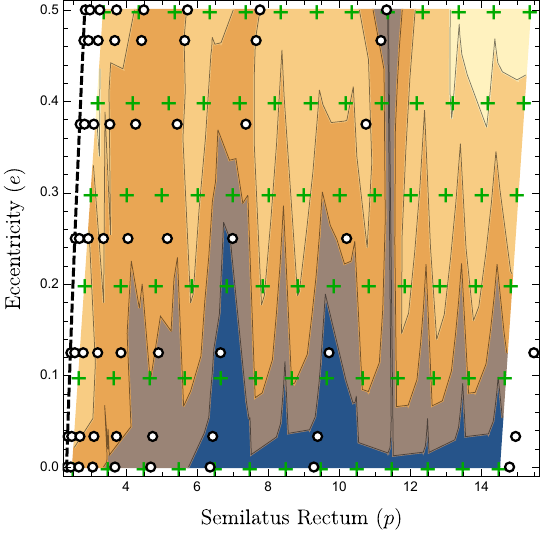}
 		\caption{$a_r$}
 		\label{fig:FrInterpolationError}
 	\end{subfigure}
 	\hfill
 	\begin{subfigure}[b]{0.7\textwidth}
 		\centering
 		\includegraphics[width=\textwidth]{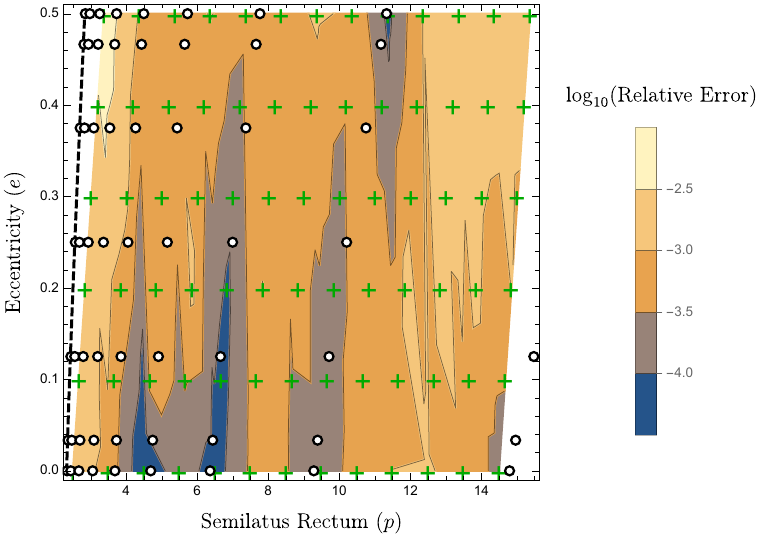}
 		\caption{$a_\phi$}
 		\label{fig:FphiInterpolationError}
 	\end{subfigure}
 	\hfill
 	\caption{The relative error of the components of the interpolated gravitational self force model for prograde equatorial orbits with $a = 0.9M$. The white dots represent the data points that were interpolated. The green crosses represent the data set that the model was tested against. The black dashed line represents the location of the last stable orbit. The relative error was calculated using the normalised $L^2$ error over a single orbital cycle.}
 	\label{fig:InterpolationError}
 \end{figure}
 
This procedure allows us to create a smooth, continuous model for the gravitational self force with relative errors less than $5 \times 10^{-3}$ in the strong field -- see Fig.~\ref{fig:InterpolationError}. 
The variation in the accuracy of the model is primarily a by-product of how close a given test point (green cross) is to the data points (white dots) used to create the model.
We note that this level of precision would not be sufficient for production grade waveforms for LISA, as we would need the relative error of the orbit averaged dissipative self-force to be less than $\sim1/\mr$, whereas the oscillatory pieces of the self-force only need to be interpolated to an accuracy of a fraction of a percent \cite{Osburn2016}. 
Our present interpolation model already likely reaches the latter criteria and a future hybrid method that combines flux and self-force data, similar to the one constructed in Ref.~\cite{Osburn2016}, can likely reach the overall accuracy goal. 
Nonetheless, our present model is more than sufficient to test our averaging procedure and to explore the effects of the GSF for eccentric Kerr inspirals. 
This will now be treated as the underlying forcing model for both the OG and NIT inspirals.

\section{Implementation} \label{section:Implementation}

Combining the above model with our action angle formulation of the osculating geodesic equations provides us with everything required to calculate the NIT equations of motion. 
We first evaluate and interpolate the various terms in the NIT equations of motion across the parameter space.
This offline process is costly but it only needs needs to be completed once. 
By contrast, the online steps are computationally cheap, which allows us to rapidly compute eccentric self-forced inspirals into a Kerr black hole.

\subsection{Offline Steps}

To make the offline calculation we complete the following steps.
\begin{enumerate}
	
\item  We start by selecting a grid to evaluate the NIT functions upon. 
We chose $y$ values between 0.2 and 0.998 in 320 equally spaced steps and $e$ values from 0.001 to 0.5 in 500 equally spaced steps (160,000 points) in the case of Kerr, or use the same spacing in $y$ but only grid in $e$ from 0.001 to 0.3 in 300 equally spaced steps (96,000 points) in Schwarzschild.\footnote{
Evaluating the NIT functions is computationally cheap so using a dense grid does not significantly increase the computational burden.
Using a dense grid also allows us to use Mathematica's default Hermite polynomial interpolation method for convenience of implementation.
The grid spacing is chosen to be sufficiently dense that interpolation error is a negligible source of error for our comparisons between the OG and NIT inspirals, though a less dense grid may also achieve this.
}

\item For each point in the parameter space $(a,y,e)$ we evaluate the functions $F_{p \backslash e}, f_{r}$ and $ s_{t \backslash\phi}$ along with their derivatives with respect to $p$ and $e$ for 30 equally spaced values of $q_r$ from $0$ to $2\pi$.

\item We then perform a fast Fourier transform on the output data to obtain the Fourier coefficients of the forcing functions and their derivatives. 

\item With these, we then use Eqs.~\eqref{eq:NIT_Relationship1}, \eqref{eq:NIT_Relationship2}, \eqref{eq:N_Operator} and \eqref{eq:NIT_Relationship3} to construct $\nit{F}_{p\backslash e}^{(1\backslash2)}$, $\nit{f}_{r}^{(1)}$ and $s_{t \backslash \phi}^{(1)}$ for that point in the parameter space.

\item We also use Eqs.~(\ref{eq:NIT_Y}) and (\ref{eq:NIT_X}) to construct the Fourier coefficients of the first order transformation functions $Y_{p\backslash e}^{(1)}$ and $X_{r}^{(1)}$.

\item We then repeat this procedure across the parameter space for each point in our grid.

\item Finally we interpolate the values for $\nit{F}_{p\backslash e}^{(1\backslash2)}, \nit{f}_{r}^{(1)}$ and $ \nit{s}_{t\backslash\phi}^{(1)}$ along with the coefficients of $Y_{p\backslash e}^{(1)}$ and $X_{r}^{(1)}$ across this grid using Hermite interpolation and store the interpolants for future use. 

\end{enumerate}
We implemented the above algorithm in Mathematica 12.2 and find, parallelized across 20 CPU cores takes, the calculation takes about one day to complete.
This is a small price to pay, since these offline steps need only be completed once.

\subsection{Online Steps}

The online steps are required for every inspiral calculation, but are comparatively inexpensive. 
The online steps for computing an NIT inspiral are as follows.
\begin{enumerate}
	
\item We load in the interpolants for $\nit{F}_{p\backslash e}^{(1\backslash2)}$ and $\nit{f}_{r}^{(1)} $ and $ \nit{s}_{ t\backslash\phi}^{(1)}$, define the NIT equations of motion.
 
 \item In order to make comparisons between NIT and OG inspirals, we also load interpolants of the Fourier coefficients of $\breve{Y}_{p/e}^{(1)}$ and $\breve{X}_{r}^{(1)}$ and Eq.~\eqref{eq:transformation} to construct first order near-identity transformations.\footnote{Note that while including $\breve{X}_{r}^{(1)}$ in the transformation is not strictly necessary, we do so anyway to further reduce the initial difference between the two inspirals.}

\item We state the initial conditions of the inspiral $(p_0, e_0, q_{r,0})$ and use the NIT to leading order in the mass ratio to transform these into initial conditions for the NIT equations of motion, i.e.,  $(\nit{p}_0, \nit{e}_0, \nit{q}_{r0})$. 

\item We then evolve the NIT equations of motion using an ODE solver (in this case Mathematica's \texttt{NDSolve}). 

\end{enumerate}

As with the offline steps we implement the online steps in Mathematica. Note that steps (ii) and (iii) are only necessary because we want to make direct comparisons between NIT and OG inspirals with the same initial conditions. In general, the difference between the NIT and OG variables will always be $\mathcal{O}(\mr)$, and so performing the NIT transformation or inverse transformation to greater than zeroth order in mass ratio will not be necessary when producing waveforms to post adiabatic order, i.e. with phases accurate to $\mathcal{O}(\mr)$.

\section{Results} \label{section:Results}

In this section we present the results from the NIT equations of motion.
We first perform some consistency checks in Sec.~\ref{section:Consistency_Checks}. 
	We then show that our NIT and OG inspirals agree to the relevant order in the mass ratio in Sec.~\ref{section:NITVsFull}. 
	Here we also compute, for the first time, self-forced inspirals in Kerr spacetime.
	With our fast NIT model we then explore the impact of the conservative effects of the first-order GSF as calculated in radiation gauge for Kerr inspirals in Sec.~\ref{section:AdbVsPA}.
	Finally, in Sec.~\ref{section:GaugeComparison}, we compare Schwarzschild inspirals calculated using a radiation gauge GSF model and a Lorenz gauge GSF model. 
\subsection{Consistency checks}\label{section:Consistency_Checks}

Before computing inspirals, we perform a series of consistency checks on the NIT equations of motion.
A useful feature of the NIT is how it separates adiabatic and post-adiabatic effects of the gravitational self-force. 
At first order in the mass ratio, this corresponds to the dissipative and conservative pieces respectively. 
We note that when we substitute  $a^\alpha \rightarrow a^\alpha_\diss$, we find that $\nit{F}_{p\backslash e}^{(2)}$, $\nit{f}_{r}^{(1)}$ and $\nit{s}_{t\backslash \phi}^{(1)}$ are numerically consistent with zero, while $\nit{F}_{p \backslash e}^{(1)}$ remains unchanged.
Similarly, when we substitute $a^\alpha \rightarrow a^\alpha_\cons$, $\nit{F}_{p \backslash e}^{(1)}$ and $\nit{F}_{p \backslash e}^{(2)}$ become consistent with zero, while $\nit{f}_{r}^{(1)}$ and $\nit{s}_{t\backslash \phi}^{(1)}$ remain the same as before. 
The functions $\nit{F}_{p\backslash e}^{(2)}$ only becomes non-zero when both dissipative and conservative effects of the first order self-force are present.

From $\nit{F}_{p \backslash e}^{(1)}$, one can calculate the average rate of change of energy and angular momentum via the following relation:
\begin{subequations}
	\begin{equation}
		\avg{\frac{d \En}{d t}} = \frac{\mr}{\Upsilon_t} \left(  \frac{\partial \En}{\partial p} \nit{F}^{(1)}_p  + \frac{\partial \En}{\partial e} \nit{F}^{(1)}_e  \right)
	\end{equation}
	\begin{equation}
		\avg{\frac{d \Lz}{d t}} = \frac{\mr}{\Upsilon_t} \left(  \frac{\partial \Lz}{\partial p} \nit{F}^{(1)}_p + \frac{\partial \Lz}{\partial e} \nit{F}^{(1)}_e \right).
	\end{equation}
\end{subequations}
We compared these to the energy and angular momentum fluxes at infinity tabulated in the Black Hole Perturbation Toolkit \cite{BHPToolkit} and generated with a variant of the \texttt{Gremlin} code \cite{Hughes2000,Hughes2001} and found that the balance laws were upheld up to relative errors $<10^{-3}$ throughout the parameter space which is consistent with the interpolation error of our self-force model.

From all of this, we can infer the significance of each of the terms in Eq.~\eqref{eq:NIT_EoM}:  $\Upsilon_r$, $\Upsilon_t$ and $\Upsilon_\phi$ capture the background geodesic motion, $\nit{F}^{(1)}_p$ and $\nit{F}^{(1)}_e$ capture the adiabatic effects due to the first order dissipative self-force, $\nit{f}^{(1)}_r$,  $\nit{s}^{(1)}_t$, and $\nit{s}^{(1)}_\phi$ capture the post-adiabatic effects due to the first order conservative self-force, and $\nit{F}^{(2)}_p$, $\nit{F}^{(2)}_e$ capture the interplay between the first order dissipative and conservative self-force, as well as the effect of the orbit averaged contribution from the second order self-force.

\subsection{Comparison between OG and NIT inspirals} \label{section:NITVsFull}

\begin{figure}
	\centering
	\includegraphics[trim={2.1cm 0 2.1cm 0},clip,width =0.9\linewidth]{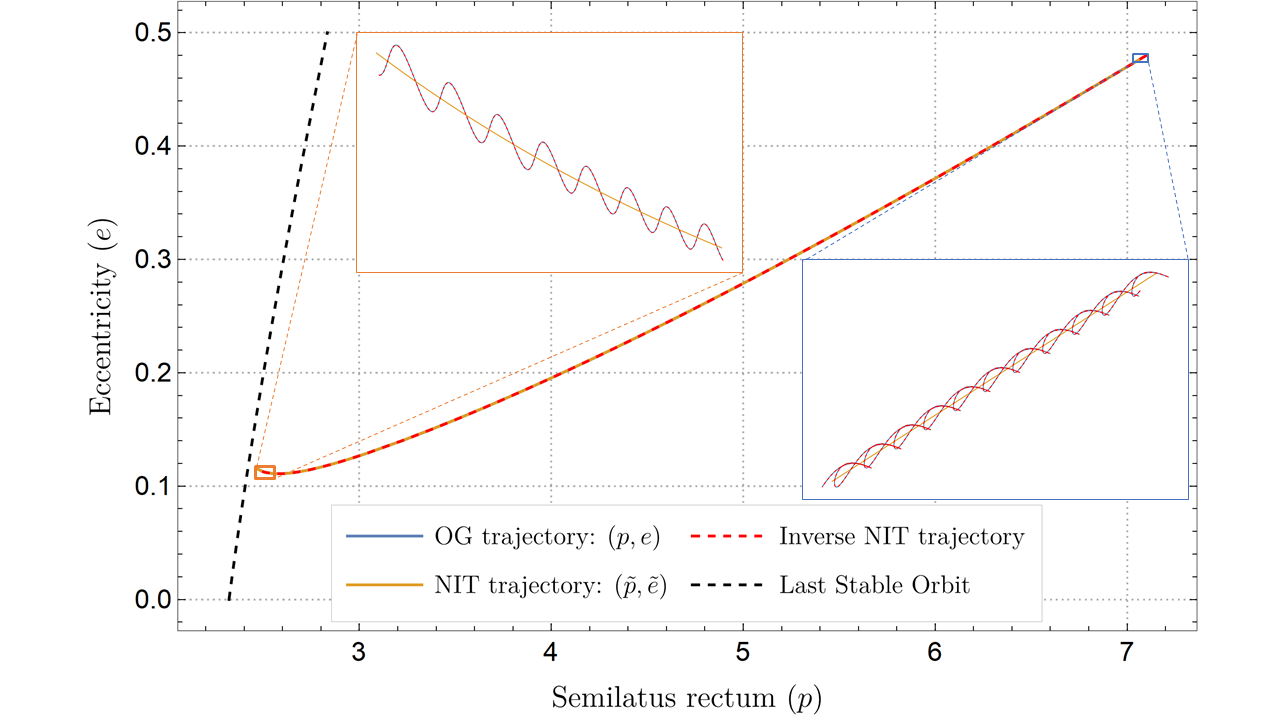}
	\caption{The trajectory through $(p,e)$ space for an inspiral with $\mr = 10^{-5}$, $a = 0.9M$, and initial conditions $(p_0 = 7.1, e_0 = 0.48)$. We show the inspiral computed using the osculating geodesic equations, the NIT equations of motion and the inverse NIT to first order in $\mr$. The insets zoom into the start and end of the inspiral to reveal the small orbital timescale oscillations. The NIT averages through these oscillations, and when using the inverse NIT to add the oscillations back on, we see that the NIT trajectory remains almost perfectly in phase with the OG trajectory throughout the inspiral.}
	\label{fig:kerrnitvsfullpeplot}
\end{figure}

In order to test the accuracy of our implementation, we compare inspirals calculated using the OG equations of motion found in Ref.~\cite{Gair2011} to those calculated using the near-identity transformed equations of motion. 
To demonstrate these results, we choose a binary with a primary of mass $M =  10^6 M_\odot$ and a secondary of mass $\mu = 10 M_\odot$ for a typical EMRI mass ratio of $\mr = 10^{-5}$. 
To push our procedure to the limit, we chose the initial conditions of our prograde inspiral to be deep in the strong field and highly eccentric with $p_0 = 7.1$ and $e_0 = 0.48$ such that the resulting inspiral would take approximately 1 year to plunge. 
We also set $q_{r,0} = t_0 = \phi_0 = 0$ for simplicity.

Figure~\ref{fig:kerrnitvsfullpeplot} shows the evolution of $p$ and $e$ over time. 
The trajectories calculated with the OG equations of motion have order $\mr$ oscillations on the orbital timescale which requires the numerical integrator to take small time steps to accurately resolve. 
The NIT trajectory does not have these oscillations so the numerical integrator can take much larger steps and still faithfully track the averaged trajectory throughout the entire inspiral.
The inverse NIT given in Eq.~\eqref{eq:inverse_trasformaiton} through $\mathcal{O}(\mr)$ can be used to add the oscillations back on to the NIT trajectory. 
We find that while this is unnecessary for computing accurate waveforms, it demonstrates that the NIT trajectory remains in phase with the OG trajectory -- see the insets of Fig.~\ref{fig:kerrnitvsfullpeplot}.

\begin{figure}
	\centering
	\includegraphics[width = 0.9\linewidth,trim={0.5cm 0 0 0},clip]{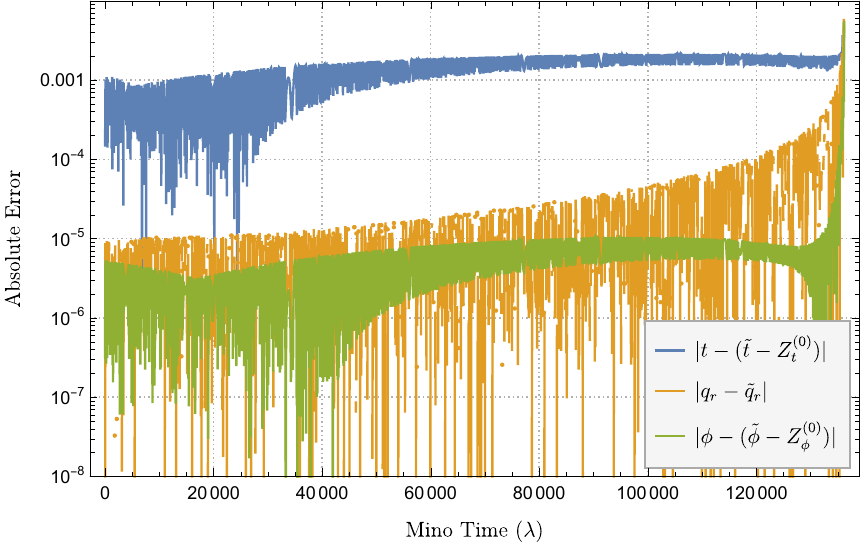}
	\caption{The difference in the orbital phase and extrinsic quantities for a equatorial Kerr inspiral with $\mr = 10^{-5}$ and $a = 0.9M$ calculated using the OG and NIT equations of motion with initial conditions $p_0 = 7.1, e_0 = 0.48$. We find that the differences remain small throughout the inspiral, only becoming large as the secondary approaches the last stable orbit where the adiabatic approximation breaks down.}
	\label{fig:phaseplot1year}
\end{figure}

The accuracy of our NIT model is further demonstrated by Fig.~\ref{fig:phaseplot1year} which shows the absolute difference in the orbital phase $q_r$ and the extrinsic quantities $t$ and  $\phi$ between the NIT and OG evolutions.
Over the course of the year long inspiral, $|t - (\nit{t} - Z_t^{(0)})| \leq 5 \times 10^{-3}$,$|\phi - (\nit{\phi} - Z_\phi^{(0)})| \leq 10^{-5}$ and $|q_r - \nit{q}_r| \leq 10^{-3}$  with the differences only spiking to $\leq 10^{-2}$ just as the trajectories reach the separatrix where the adiabatic approximation breaks down. 

\begin{figure}[pt]
	\centering
	\begin{subfigure}[b]{0.9\linewidth}
		\centering
		\includegraphics[width=0.9\textwidth]{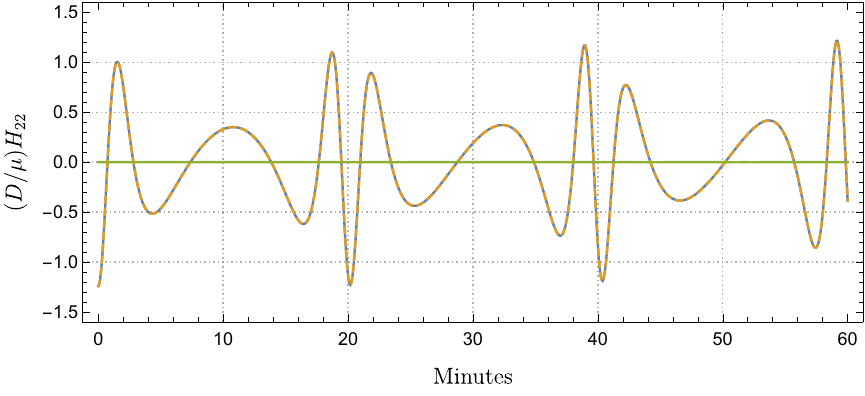}
		\caption{First hour of the waveform.}
		\label{fig:EarlyWF}
	\end{subfigure}
	\hfill
	\begin{subfigure}[b]{0.9\textwidth}
		\centering
		\includegraphics[width=0.9\textwidth]{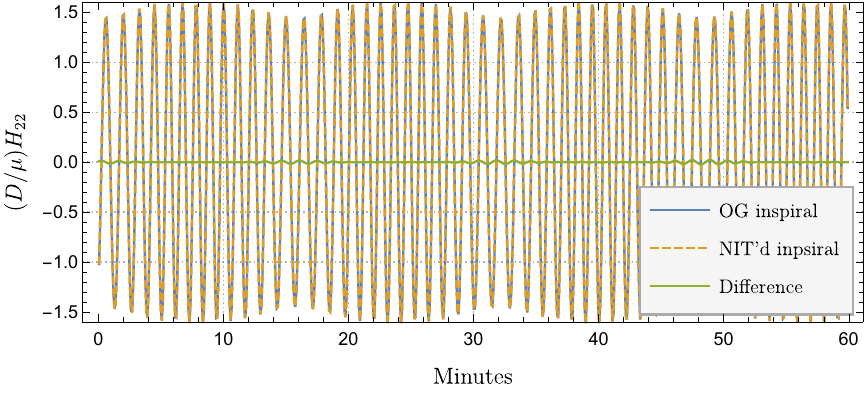}
		\caption{Last hour of the waveform.}
		\label{fig:LateWF}
	\end{subfigure}
	\hfill
	
	\caption{Two snapshots of the dominant $(l,m) = (2,2)$ mode of the quadrupole waveform for our prograde, equatorial Kerr inspiral with $(a, \mr, p_0,e_0) = (0.9M, 10^{-5}, 7.1,0.48)$. These snapshots correspond to the first and last hours of the inspiral. This shows that the waveform generated using the NIT trajectory almost perfectly overlaps with the waveform generated using the OG trajectory. It also demonstrates how dramatically an EMRI waveform evolves throughout the inspiral.}
	\label{fig:KerrWFPlots}
\end{figure}

Finally, we test the effect the NIT procedure has on the waveform. 
In principle, we could use our averaged equations of motion in conjunction with the \texttt{FastEMRIWaveforms} (FEW) framework to rapidly compute waveforms with relativistic amplitudes. 
However, currently, the FEW framework only has amplitude data for Schwarzschild inspirals.
As such, we make use of the same procedure as the Numerical Kludge \cite{Babak2007} by mapping the Boyer-Lindquist coordinates $\{t,r,\theta,\phi\}$ to flat space coordinates and using the  quadrupole formula to generate the waveform. 
The resulting waveforms are only an approximation to the true waveforms, but since both inspiral trajectories are being fed through the same waveform generation scheme this should not bias the results when finding the difference in the waveform as a result of using the NIT trajectory instead of the OG trajectory.

\begin{figure}
	\centering
	\includegraphics[width=0.9\linewidth]{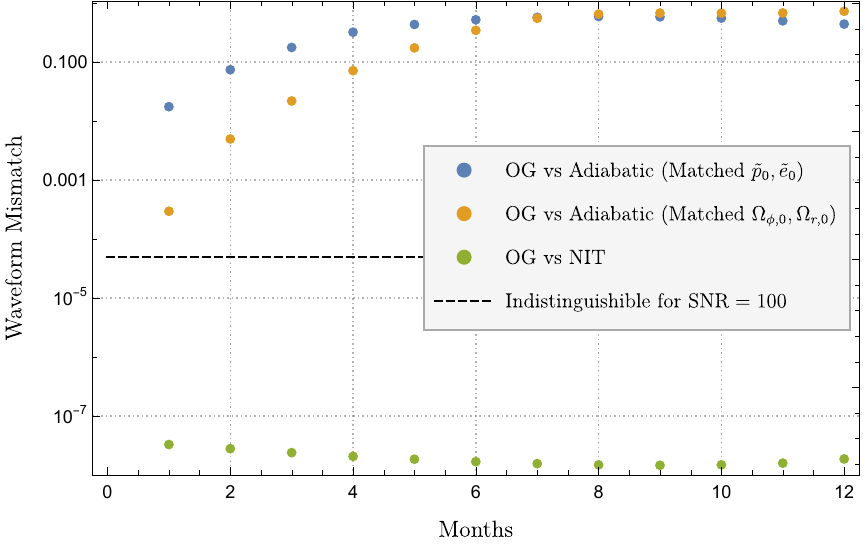}
	\caption{The mismatch between the semi-relativistic quadrupole waveforms between inspirals calculated using the OG equations with $(a, \mr, p_0,e_0) = (0.9M, 10^{-5}, 7.1,0.48)$ and the adiabatic EOM matched initial conditions, the adiabatic EOM calculated with matched initial frequencies, and the near-identity transformed EOM. We also mark the mismatch that would be indistinguishable for signals with SNR $= 100$.}
	\label{fig:WFMismatch}
\end{figure}

From Fig.~\ref{fig:KerrWFPlots}, we can see that the waveforms generated by each evolution scheme,sampled every $t = 1M \approx 5s$, are almost identical by eye.
We can further quantify this by calculating the waveform mismatch using the \texttt{WaveformMatch} function from the \texttt{SimulationTools}~\cite{SimulationTools} Mathematica package and assuming a flat noise curve. 
From Fig.~\ref{fig:WFMismatch}, we see that the mismatch remains below $5 \times 10^{-8}$ throughout the inspiral. 
At this level of mismatch the two waveforms would be completely indistinguishable for EMRIs with  signal-to-noise ratio (SNR) of upto (at least) 3000  \cite{Flanagan1997, Lindblom2008, McWilliams2010}. 

\begin{figure}
	\centering
	\includegraphics[width=0.9\linewidth]{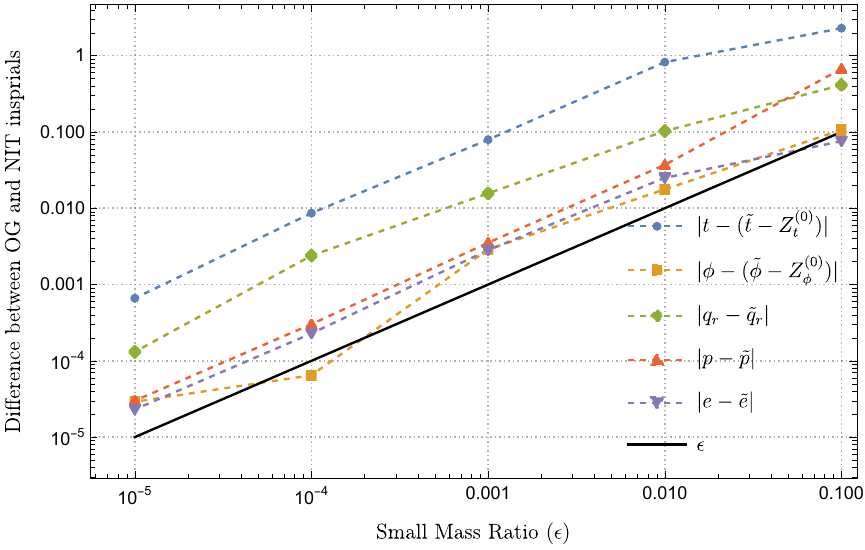}
	\caption{The absolute difference in the quantities of a prograde inspiral with $a=0.9M$ and $e_0 = 0.2$ after evolving from $p = 4$ to $p = 3$ using either the OG or NIT equations of motion. We observe that all the differences follow the solid, black $\mr$ reference curve, as expected. }
	\label{fig:equatorialchebyshevabsoluteerror}
\end{figure}

Next, the difference between the OG and NIT quantities should scale linearly with the mass ratio. 
This is illustrated in Fig.~\ref{fig:equatorialchebyshevabsoluteerror}, where starting with initial conditions $p_0 = 4$ and $e_0 = 0.2$ we evolved the inspiral until it reached $p = 3$ for mass ratios ranging from $10^{-1}$ to $10^{-5}$. 
While working with only machine precision arithmetic we found that for smaller mass ratios the numerical error of the solver of the OG inspiral became dominant over the difference with the NIT. 
To rectify this, we increased the working precision of our solver to 30 significant digits and found that the difference does, in fact, scale linearly with the mass ratio. 
This requirement for higher precision only  affected the OG solver, the NIT equations of motion can be solved with machine precision arithmetic without introducing any significant error.

\begin{figure}
	\centering
	\includegraphics[width=0.9\linewidth, clip]{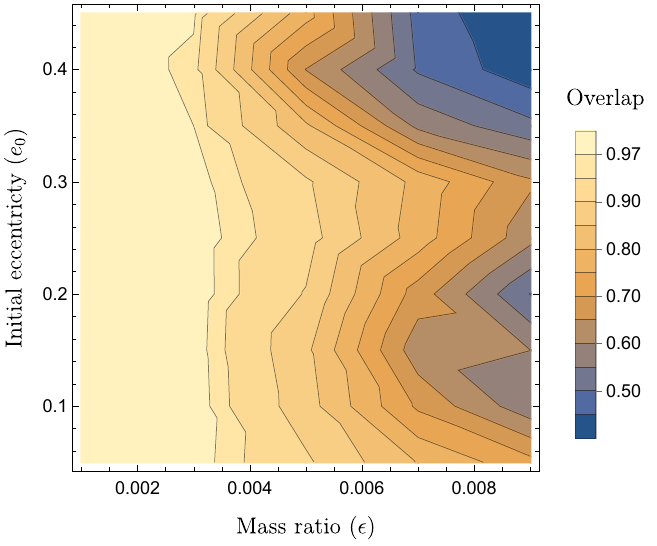}
	\caption{The overlap between OG and NIT waveforms for year-long, prograde, $a=0.9M$, equatorial Kerr inspirals as a function of the mass ratio and initial eccentricity. The difference between the two waveforms is less than the accuracy benchmark of 0.97 for mass ratios $ \leq 3 \times 10^{-3}$, but not for mass ratios larger than this. While increasing eccentricity does have an effect on the overlap, this effect is not as strong as the effect observed in Fig.~9 of Ref.~\cite{McCart2021}.}
	\label{fig:EquatorialKerrWFMismatch}
\end{figure}

Since the difference between OG and NIT quantities scales with the mass ratio, it is natural to ask how large can the mass ratio be before the NIT and OG waveforms differ enough to affect data analysis.
Following the procedure outlined in Ref.~\cite{McCart2021}, we used our fast NIT inspiral code along with a root-finding algorithm to find the initial value of $p$ that corresponds to a year long inspiral for a given value of the mass ratio and initial eccentricity, and assuming a primary mass of  $10^6 M_{\odot}$. 
We use these initial conditions to calculate the overlap between year-long NIT and OG waveforms. 
This calculation is repeated with mass ratios $\mr = \{1, 3, 5, 7, 9 \} \times 10^{-3}$ and initial eccentricities $e_0$ ranging from $0.05$ to $0.45$ in equally spaced steps of $0.05$. 
The result of this analysis can be seen in Fig.~\ref{fig:EquatorialKerrWFMismatch}. 
This demonstrates that NIT and OG waveforms have overlaps larger than the benchmark of 0.97 \cite{Owen1995} for mass ratios less than $ \approx 3 \times 10^{-3}$, but these overlaps decrease substantially for mass ratios larger than this. 
We also see that the overlap generally decreases as the initial eccentricity increases, though this effect is not as strong as the effect demonstrated by a similar analysis in Ref.~\cite{McCart2021} for NITs applied to highly eccentric inspirals in Schwarzschild. 
They also found that the mismatch between  NIT and OG waveforms became substantial for mass ratios larger than $2 \times 10^{-4}$.
These differences between the two analyses are most likely the result of our inspirals being deeper in the strong field and driven by a self-force computed in a different gauge (Ref.~\cite{McCart2021} uses the Lorenz gauge self-force).
Such mismatches should not be an issue for EMRI data analysis as EMRIs have mass ratios that range from $10^{-7}$ to $10^{-4}$. 
However, these mismatches become significant for intermediate mass ratio inspirals, with mass ratios between $10^{-4}$ to $10^{-1}$.
Since both the OG and NIT equations of motion are formally valid to the same order in the mass ratio, it is not clear a priori which of the two would be closer to the true inspiral. 
When completed at 1-post-adiabtatic (1PA) order the two sets of equations represent different resummations of the 1PA equations of motion, differing only in their higher order (2+) PA terms.
The fact that we are seeing a significant difference between these two resummations for intermediate mass ratios suggests that such higher order PA terms might become relevant. 
However, in this case it might just be signalling the importance of the missing orbit-averaged dissipative self-force term at 1PA order.

\begin{table}
\begin{center}
	\begin{tabular}{c | c c c} 
		\hline
		$\mr$ & OG Inspiral & NIT Inspiral & Speed-up  \\ [0.5ex] 
		\hline
		$10^{-2}$ & 44s & 0.85s & $ \sim 37$ \\ 
		$10^{-3}$ & 6m 48s & 0.78s & $\sim 491$ \\
		$10^{-4}$ & 54m 12s & 0.81s & $\sim 3782$ \\
		$10^{-5}$ & 6hrs 16m & 0.76s & $\sim 29655$ \\ 
		\hline 
	\end{tabular}
\end{center}
\caption{Computational time required to evolve an inspiral from its initial conditions of $p_0 = 7.1$ and $e_0 = 0.48$ to the last stable orbit for different values of the mass ratio, as calculated in Mathematica 12.2 on an Intel Core i7 @ 2.2GHz. The computational time for the OG inspiral scales inversely with the mass ratio, whereas the computational time for NIT inspirals is independent of the mass ratio. This demonstrates how the smaller the mass ratio of the inspiral, the greater speed-up one  obtains from using the NIT equations of motion.\label{table:computation_time}}
\end{table}

Finally, we note that using the NIT equations of motion produces a substantial speed-up over using the OG equations.
From Table~\ref{table:computation_time}, we see the typical computation time for an inspiral starting at $p_0 = 7.1$ and $e_0 = 0.48$ and evolved until the inspiral reaches the last stable orbit for different values of the mass ratio. 
We see that as we decrease the mass ratio by an order of magnitude, the OG inspiral takes roughly an order of magnitude longer to compute, as it would have to resolve an order of magnitude more orbital cycles before reaching last stable orbit.
The NIT inspirals all take roughly the same amount of time to evolve to the last stable orbit, regardless of the mass ratio. 
Using our current Mathematica implementation, the NIT inspirals can be computed in less than a second.
This time could be further reduced tens of milliseconds if one uses a compiled language such as C/C++, as was done in Paper I \cite{NITs}.
We see that using the NIT equations of motion is most advantageous for long inspirals with small mass ratios.
Another benefit of using the NIT is that the inspiral requires taking fewer time steps, which results in less numerical error, making it easier achieve a given target accuracy.

The only disadvantage of our formulation is that our final trajectory is parametrized in terms Mino time $\lambda$, whereas LISA data analysis applications will need waveforms parametrized Boyer-Lindquist retarded time $t$. 
Since our formulation also outputs $t(\lambda)$, we can numerically invert this to get $\lambda(t)$ which allows us to resolve this issue at the cost of additional computation time. 
This was also a problem with the NIT formulation in Schwarzschild where the final trajectory is outputted as a function of the quasi-Keplerian angle $\chi$ \cite{NITs,McCart2021}. 
This problem might be circumvented entirely by performing an additional transformation to our NIT equations of motion which would produce averaged equations of motion parametrized by $t$ as outlined in \cite{Pound2021}. 

Since we are now satisfied that our formulation can produce fast and accurate self-force driven trajectories, we can now use this procedure to explore the phenomenology of eccentric, equatorial Kerr inspirals.

\subsection{Impact of adiabatic and post-adiabatic effects} \label{section:AdbVsPA}

\begin{figure}
	\centering
	\begin{subfigure}[b]{0.496\textwidth}
		\centering
		\includegraphics[width=\textwidth]{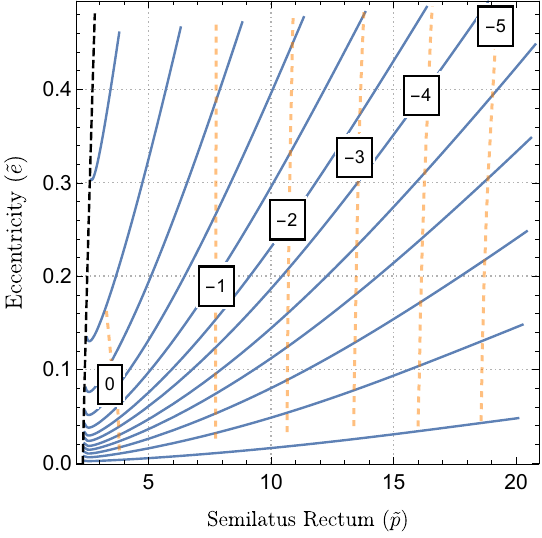}
		\caption{Prograde inspirals}
		\label{fig:progradeKerrPS}
	\end{subfigure}
	\hfill
	\begin{subfigure}[b]{0.496\textwidth}
		\centering
		\includegraphics[width=\textwidth]{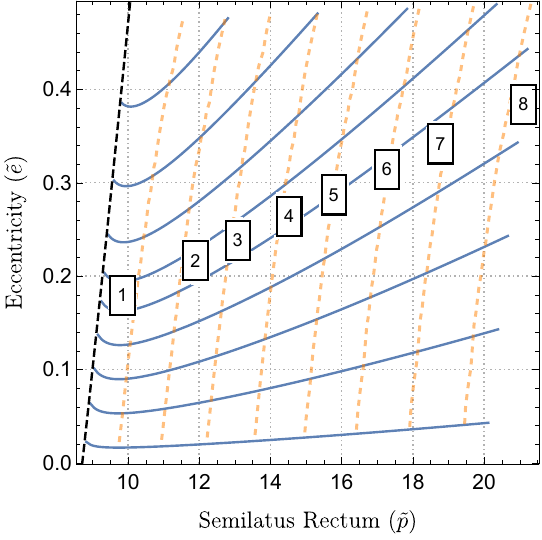}
		\caption{Retrograde inspirals}
		\label{fig:retrogradeKerrPS}
	\end{subfigure}
	\caption{Sample trajectories through $(\nit{p},\nit{e})$ space for prograde and retrograde equatorial Kerr inspirals with $\mr = 10^{-5}$ and $a = 0.9 M$. 
	From these plots, we see the familiar behaviour of EMRIs losing eccentricity as the compact object approaches the primary  and then gaining eccentricity just before crossing the separatrix (dashed black line). 
	The dashed orange curves are contours that mark the number of radians $\nit{q}_{r,0}$ will evolve from a given point until plunge. 
	The conservative self-force for retrograde orbits has a similar effect to the non-spinning case as it causes $\nit{q}_{r,0}$ to increase throughout the inspiral. 
	In the prograde case, $\nit{q}_{r,0}$ decreases for most of the inspiral and then slightly increases shortly before plunge.}
	\label{fig:Kerr_Inspirals}
\end{figure}

With the ability to generate fast and accurate inspirals,  we can survey the physics of equatorial Kerr inspirals and examine how this differs from the Schwarzschild case. 
From Fig.~\ref{fig:progradeKerrPS}, we see the familiar effect of gravitational radiation reaction on the semilatus rectum, $\nit{p}$, and eccentricity, $\nit{e}$, whereby $\nit{p}$ and $\nit{e}$ both decrease over the inspiral with $\nit{e}$ growing a little as the last stable orbit is approached \cite{Glampedakis2002,Fujita:2020zxe,Isoyama:2021jjd}.
As the inspiral approaches the last stable orbit adiabaticity breaks down and the inspiral undergoes a transition to plunge \cite{Thorne00, Burke2019, Compere21,Sundararajan2008}. 
As such, we stop our inspirals just before the last stable orbit.
Our results are the first inspirals to include conservative self-force corrections to the equations of motion in Kerr spacetime.
The initial phase $q_{r,0}$ only evolves secularly when conservative self-force corrections are present and so we use this as a measure of the influence of these corrections \cite{Warburton2012}.
This is illustrated by the dashed orange curves in Fig.~\ref{fig:progradeKerrPS}, which mark the number of radians $\nit{q}_{r,0}$ will evolve from a given pair of initial conditions $(\nit{p}_0,\nit{e}_0)$ until the last stable orbit.
For retrograde Kerr (and Schwarzschild orbits in Fig.~\ref{fig:Schwarzschild_Inspirals}), we find that $\nit{q}_{r,0}$ increases throughout the inspiral, whereas for prograde Kerr  $\nit{q}_{r,0}$ decreases  during the inspiral before increasing slightly just before plunge. 
This is consistent with the change of sign in the correction to the rate of periapsis advance induced by the conservative self force as a function of spin in the circular orbit limit \cite{VanDeMeent2017} -- see \ref{Periastron_Advance} for further details.

\begin{figure}
	\centering
	\includegraphics[width=0.9\linewidth]{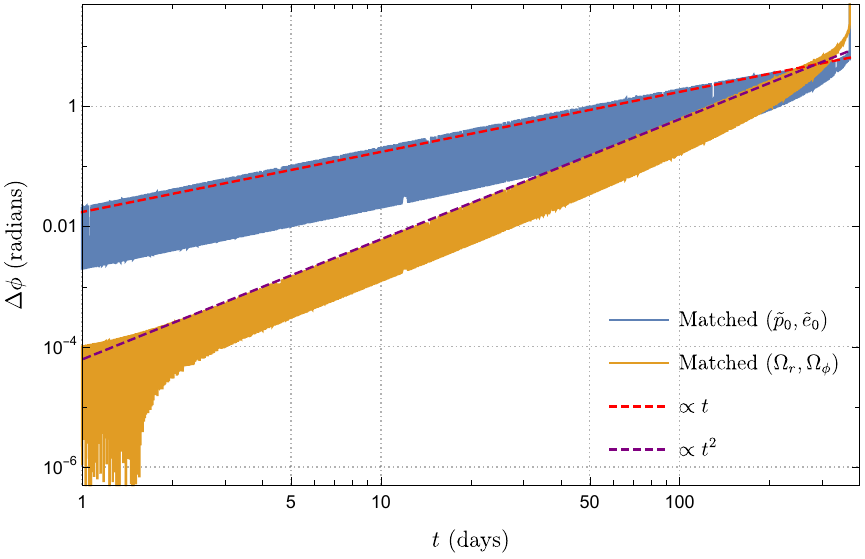}
	\caption{Difference in $\phi$ as a function of $t$ between an adiabatic and a first order self-forced inspiral when either matching initial conditions or matching the initial Boyer-Lindquist frequencies. 
	The self-forced inspiral has initial conditions $(\nit{p}_0, \nit{e}_0) = (7.1, 0.48)$ with mass ratio $\mr = 10^{-5}$. 
	Matching initial conditions results in an error that grows linearly with $t$, while matching frequencies produces an error that is initially constant and then grows quadratically with $t$.}
	\label{fig:MatchFreqvsICsKerrEq}
\end{figure}

As discussed in Sec.~\ref{section:Consistency_Checks}, one can readily calculate adiabatic inspirals using the NIT equations of motion by simply neglecting the post-adiabatic terms. 
However, when trying to determine  how post-adiabatic corrections effect the inspiral, one must be mindful of how one matches up an adiabatic inspiral with its post-adiabatic counterpart. 
Following the argument found in Refs.~\cite{Warburton2012} and \cite{Osburn2016}, matching the initial conditions $(\nit{p}_0, \nit{e}_0)$ results in an error in the orbital phases that grows linearly in $t$ as the conservative self-force changes the orbital frequencies \cite{Barack2011}.
Instead, one should instead match the Boyer-Lindquist time fundamental frequencies $\Omega_r$ and $\Omega_\phi$. 
For an adiabatic inspiral, these are directly related to the Mino-time fundamental frequencies via $\Omega^{\Ad}_{r \backslash \phi} = \frac{\Upsilon_{r \backslash \phi}}{\Upsilon_t}$ \cite{Fujita2009a}.
To calculate these frequencies as perturbed by the conservative self-force, one can either follow the method outlined in Ref.~\cite{Osburn2016}, or one can calculate them directly from the NIT equations of motion:
\begin{equation}\label{eq:NIT_Frequencies}
	\Omega^{\PA}_{r} = \frac{\Upsilon_{r} + \mr \nit{f}^{(1)}_{r}}{\Upsilon_t + \mr \nit{s}^{(1)}_t} +  \HOT{2} 
	\quad \text{and} \quad
	\Omega^{\PA}_{\phi} = \frac{\Upsilon_{\phi} + \mr \nit{s}^{(1)}_{ \phi}}{\Upsilon_t + \mr \nit{s}^{(1)}_t} + \HOT{2}.
\end{equation}
We find that both approaches give the same result up to an error that scales as $\mr^2$. 
With this in hand, we can now choose a value for our initial conditions $(\nit{p}^{\PA}_0, \nit{e}^{\PA}_0)$ for our self-forced inspiral, and then root find for initial conditions $(\nit{p}^{\Ad}_0, \nit{e}^{\Ad}_0)$ that satisfy the simultaneous equations
 \begin{subequations} \label{eq:Adiabatic_Frequency_Match}
 	\begin{equation}
 		\Omega^{\PA}_r(\nit{p}^{\PA}_0, \nit{e}^{\PA}_0) -\Omega^{\Ad}_r(\nit{p}^{\Ad}_0, \nit{e}^{\Ad}_0) = 0, 
 	\end{equation}
 	\begin{equation}  
 		\Omega^{\PA}_\phi(\nit{p}^{\PA}_0, \nit{e}^{\PA}_0) -\Omega^{\Ad}_\phi(\nit{p}^{\Ad}_0, \nit{e}^{\Ad}_0) = 0. 
 	\end{equation}
 \end{subequations}

Using this procedure to match the initial frequencies we find that the linear-in-$t$ growth of the difference in the orbital phases is removed and the phase difference grows quadratically in $t$ as expected -- see Fig.~\ref{fig:MatchFreqvsICsKerrEq}. 

\subsection{Comparing inspirals driven using radiation Gauge and Lorenz gauge self-force in Schwarzschild spacetime} \label{section:GaugeComparison}
We now turn our attention to the special case of Schwarzschild ($a = 0$), where we now have interpolated GSF models calculated in two different gauges. 
In addition to our outgoing radiation gauge self-force model, we make use of an interpolated Lorenz gauge self-force from Ref.~\cite{Warburton2012}, which is valid in the domain $6 \leq p \leq 12$ and $0 \leq e \leq 0.2$. 
We apply the same NIT procedure to inspirals driven by this force model, and find agreement with inspirals calculated in Paper I, up to the precision of the numerical solver.

\begin{figure}
	\centering
	\begin{subfigure}[b]{0.496\textwidth}
		\centering
		\includegraphics[width=\textwidth]{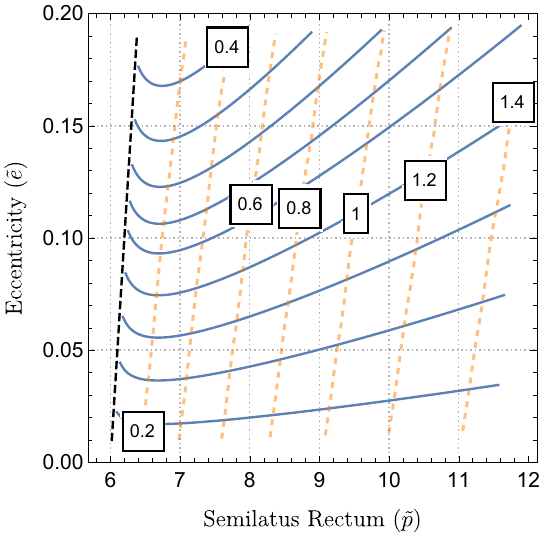}
		\caption{Outgoing radiation gauge inspirals}
		\label{fig:RadiationGaugeInspirals}
	\end{subfigure}
	\hfill
	\begin{subfigure}[b]{0.496\textwidth}
		\centering
		\includegraphics[width=\textwidth]{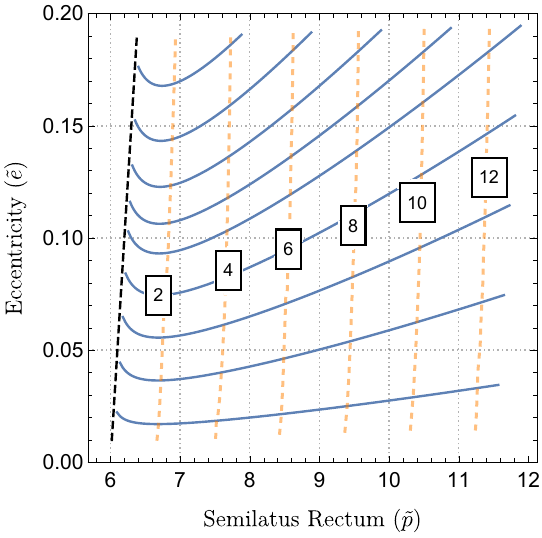}
		\caption{Lorenz gauge inspirals}
		\label{fig:LorenznGaugeInspirals}
	\end{subfigure}
	\hfill
	
	\caption{Sample Schwarzschild trajectories through $(\nit{p},\nit{e})$ space using either a radiation gauge or a Lorenz gauge model, accompanied by contours denoting the change in $\nit{q}_{r,0}$ (in radians) by the end of the inspiral if the inspiral had started in that point of the parameter space. 
	While there are only slight differences in the $(\nit{p},\nit{e})$ trajectories, there is a stark difference in the evolution of $\nit{q}_{r,0}$ induced by each model.}
	\label{fig:Schwarzschild_Inspirals}
\end{figure}
To assess the accuracy of the dissipative self-force, we calculate the orbit averaged energy and angular momentum fluxes, and find that they agree with values from the literature with a relative error less than $10^{-3}$ for both models across the parameter space.
To assess the accuracy of the conservative self-force, we calculate the periapsis advance in the circular orbit limit as outlined in \cite{Barack2010a} using the formula found in \cite{VanDeMeent2017}. 
We find that both models show good agreement with the literature across the Lorenz gauge model's domain of validity, with the Lorenz gauge model producing errors less than $10^{-3}$ and the radiation gauge model producing relative errors less than $10^{-2}$.

While we find good agreement between the two results for gauge invariant quantities, we see from Fig.~\ref{fig:Schwarzschild_Inspirals} that the inspirals experience dramatically different conservative effects, depending on the gauge used. 
While in both cases, the conservative self-force acts against geodesic periapsis advance, we see that the evolution of $q_{r,0}$ depends heavily on the gauge involved, while the trajectories through $p$ and $e$ space are less affected. 
This is to be expected as the leading order averaged rates of change of $p$ and $e$ are related to the gauge invariant asymptotic fluxes, while the change in $q_{r,0}$ is induced entirely by the (gauge dependent) conservative self-force \cite{VandeMeent2016}. 

Just as when comparing adiabatic and self-forced inspirals it is important to match the initial frequencies (rather than the initial $(p,e)$ values). 
We note that for the Lorenz gauge model, we must account for the fact that the perturbed time coordinate, $\hat{t}$, is not asymptotically flat \cite{Sago:2008id}.
We can define an asymptotically flat time coordinate for Lorenz gauge inspirals via the following rescaling
\begin{equation}\label{eq:Lorenz_t_rescale}
	t = (1 + \mr \alpha) \hat{t}.
\end{equation}
where $\alpha$ is given by
\begin{equation}
	\alpha(p,e) = -\frac{1}{2} h_{tt}^{(1)} (r \rightarrow \infty).
\end{equation}
We make use of a code provided to us by S.~Akcay to numerically calculate this quantity for Lorenz gauge values of $p$ and $e$ \cite{Akcay2015,Akcay2013}.
Equation \eqref{eq:Lorenz_t_rescale} means the perturbed Boyer-Lindquist frequencies must also be rescaled by:
\begin{equation}
	\Omega_{r}^{(LG)}  = (1 - \mr \alpha)\frac{\Upsilon_{r} + \mr \nit{f}^{(1)}_{{r} (LG)}}{\Upsilon_t + \mr \nit{s}^{(1)}_{t (LG)}}
	\quad \text{and} \quad 
	\Omega_{\phi}^{(LG)} = (1 - \mr \alpha)\frac{\Upsilon_{\phi} + \mr \nit{s}^{(1)}_{{\phi} (LG)}}{\Upsilon_t + \mr \nit{s}^{(1)}_{t (LG)}}
	.
\end{equation}
In the radiation gauge model, the corresponding subtleties have been dealt with by including the gauge completion corrections, so the frequencies can be calculated using Eq.~\eqref{eq:NIT_Frequencies} as before. 
Thus, we can choose a value for $\nit{p}_{0}^{(LG)}$ and $\nit{e}_{0}^{(LG)}$ in Lorenz gauge and root find for values of  $\nit{p}_{0}^{(RG)}$ and $\nit{e}_{0}^{(RG)}$ in radiation gauge that satisfy: 
\begin{subequations}
	\begin{equation}
		\Omega_r ^{(RG)} (\nit{p}_{0}^{(RG)},\nit{e}_{0}^{(RG)} ) - \Omega_r ^{(LG)}(\nit{p}_{0}^{(LG)},\nit{e}_{0}^{(LG)} ) = 0,
	\end{equation}
	\begin{equation}
		\Omega_\phi ^{(RG)} (\nit{p}_{0}^{(RG)},\nit{e}_{0}^{(RG)} ) - \Omega_\phi ^{(LG)}(\nit{p}_{0}^{(LG)},\nit{e}_{0}^{(LG)} ) = 0.
	\end{equation}
\end{subequations}
This allows us to make comparisons between inspirals driven by self-force models calculated in different gauges. 
We use an inspiral driven by the Lorenz-gauge force model with initial conditions $(p_0,e_0) = (11, 0.18)$, mass ratio $\mr = 10^{-5}$ as our reference inspiral which should last just over two and a half years for a $10^6 M_\odot$ primary.

\begin{figure}
	\centering
	\includegraphics[width=0.9\linewidth]{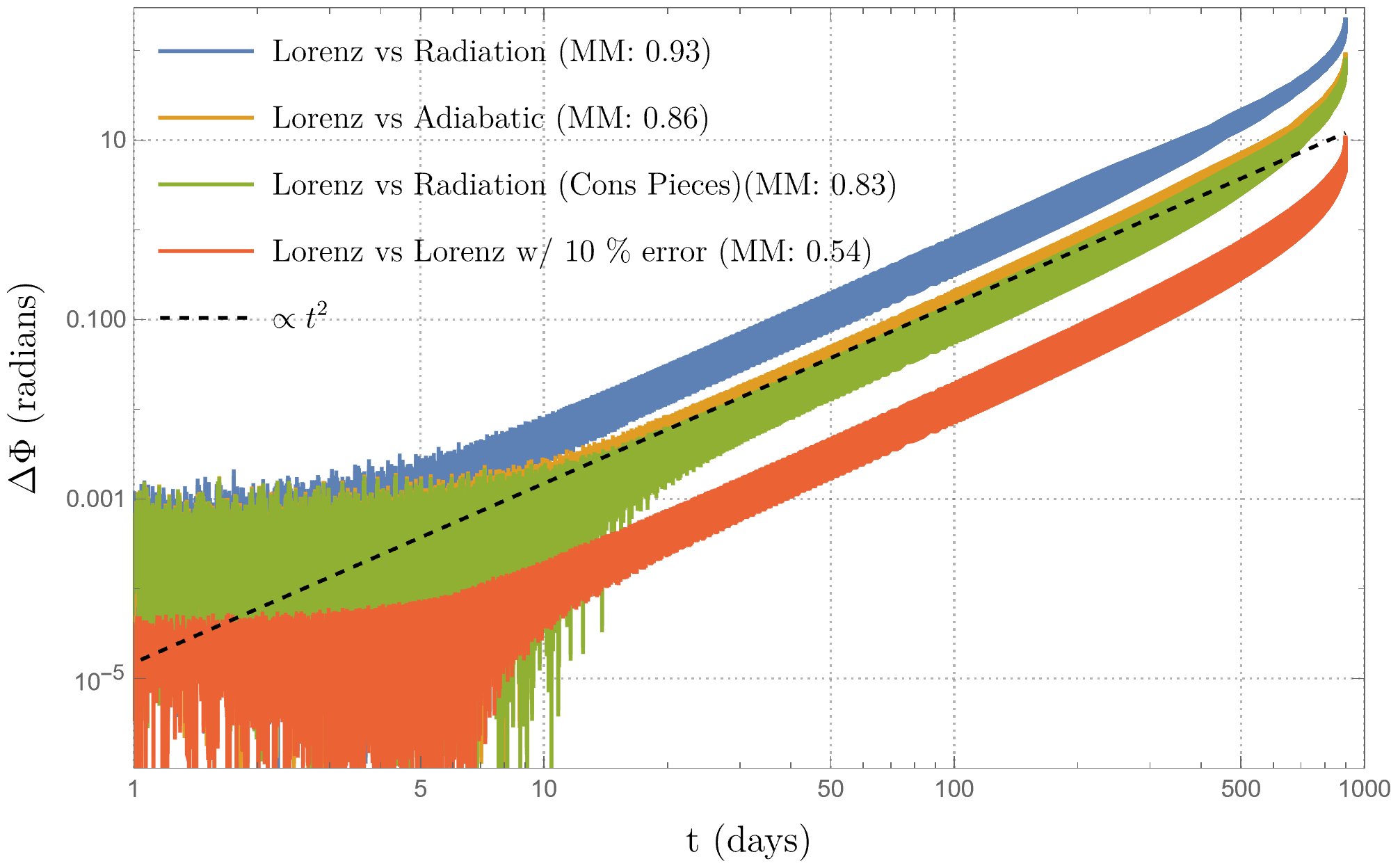}
	\caption{The difference in the waveform phase $\Phi$ for various inspirals as a function of $t$ when compared to NIT inspiral driven by a Lorenz gauge self-force model, with initial conditions $(a, p_0, e_0) = (0, 11, 0.18)$, mass ratio $\mr = 10^{-5}$, viewing angles $\Theta = \pi/4$ and $\varPhi = 0$, and sampled every $\Delta t = 1M \sim 5s$. We also show the mismatch (MM) between the waveforms in each case. By matching the initial frequencies, we compare an inspiral calculated using a radiation gauge self-force model, an adiabatic inspiral, an inspiral with the adiabatic pieces of the Lorenz gauge model and conservative pieces from the radiation gauge model, and a Lorenz gauge model with a 10 \% relative error added to each conservative piece. In all cases the difference grows quadratically in time. This plot suggests that post-adiabatic waveforms calculated using only the first-order self-force differ significantly depending on the gauge used.}
	\label{fig:Gauge_Comparison}
\end{figure}

In Fig.~\ref{fig:Gauge_Comparison}, we see the difference in the phase of the waveform $\Phi$ as a function of time between the Lorenz gauge NIT inspiral, and a number of reference models. 
We make use of the relations between the NIT quantities and the waveform phases derived in Ref.~\cite{McCart2021} to find
\begin{equation}
	\Phi_{r} =  \nit{q}_r - \Omega_r^{\PA} Z^{(0)}_t + \mathcal{O}(\mr) 
	\quad
	\text{ and}
	\quad
	\Phi_{\phi} =  \nit{\phi} - \Omega_\phi^{\PA} Z^{(0)}_t + \mathcal{O}(\mr).
\end{equation}

We then feed the solutions for $\{\nit{p}(t), \nit{e}(t),\Phi_{r}(t),\Phi_{\phi}(t)\}$ into the FastEMRIWaveforms package to generate these eccentric Schwarzschild waveforms \cite{Chua2021a}. 
Finally, we make use of the $\texttt{SimulationTools}$ Mathematica package to calculate the mismatches and decompose the waveforms into a single evolving amplitude $A(t)$ and phase $\Phi(t)$. This allows us to find the difference in the waveform phase $\Delta \Phi(t)$ between the Lorenz gauge inspiral and the other inspiral calculations. We use this as our point of comparison as the waveform phase is an observable and thus a gauge invariant quantity.

We note that in each case, we see constant error which gives way to quadratic growth with $t$ just as in Fig.~\ref{fig:MatchFreqvsICsKerrEq}. 
As we discussed in Sec.~\ref{section:AdbVsPA}, this shows that the initial frequencies were correctly matched.
From the blue curve, we see that the NIT radiation gauge inspiral quickly goes out of phase with the Lorenz gauge NIT inspiral, resulting in a very large mismatch of 0.93. 
We found that the largest source of error here is due to interpolation error for in the adiabatic pieces of the NIT. 
Since these are related to the gauge invariant fluxes, these pieces should be identical in both models. 
As such, we can rectify this error by using the Lorenz gauge functions for the adiabatic pieces and continue to use the radiation gauge functions for the conservative pieces of the NIT equations of motion.
The improvement is evident in the green curve, which shows much better agreement with the Lorenz gauge NIT inspiral, with the mismatch falling to 0.83. 
However, it is only slightly better than matching an adiabatic inspiral (orange curve) using Eq.~\eqref{eq:Adiabatic_Frequency_Match} resulting in a mismatch of 0.86. Both radiation gauge and adiabatic inspirals go out of phase by almost 100 radians by the time they reach the last stable orbit. 

In order to rule out the possibility of interpolation error of the conservative effects being the primary cause of this difference, we repeat the Lorenz gauge inspiral, but this time we manually add a relative error of $\delta = 0.1$ to all of the conservative pieces of both the 
NIT equation of motion and our matching procedure for the initial conditions, e.g., $ \dot{\nit{q_r}} = \Upsilon_r + \mr \nit{f}_r^{(1)} \rightarrow \Upsilon_r + \mr \left(\nit{f}_r^{(1)} + \delta \nit{f}_r^{(1)} \right) $ etc.
We note that this is an order of magnitude larger than the $10^{-2}$ error produced by the radiation gauge model when calculating the gauge invariant quasi-circular periapsis advance. 
From the red curve, we see that manually adding a constant 10\% relative error results in phase difference and a mismatch (0.54) which is significantly smaller than what we observe between the two self-forced inspirals. This gives us confidence that this difference is not dominated by numerical error.

From these investigations, we infer that the trajectories driven using only the first order self-force are gauge dependent, and thus, so too are their waveforms. 
Since post-adiabatic waveforms are an observable quantity, this leads us to conclude that incorporating the orbit-averaged dissipative second-order self-force will be necessary to obtain gauge invariant, post-adiabatic waveforms. 
Moreover, since the difference between the Radiation and Lorenz gauge self-forced inspirals is of the same magnitude as the difference with the adiabatic inspiral, we further conclude that the impact of the orbit-averaged dissipative second-order self-force must be of a similar magnitude in at least one of the two gauges.

\section{Discussion} \label{section:Discussion}

%
In this paper, we present the first self-forced inspirals in Kerr spacetime. 
We computed the self-force in the radiation gauge using the code of Ref.~\cite{VandeMeent2016} and interpolated it over a region of the parameter space of eccentric, equatorial orbits using Chebyshev interpolation.
Our model achieves sub-percent accuracy for the self-force across the two dimensional parameter space using only 105 points is a substantial improvement over cubic spline interpolation which would require $ \mathcal{O}(10^3)$ points to achieve a comparable level of accuracy.
So far we have applied our method to strong-field regions of the parameter space for three values of the primary's spin ($a=0,\pm0.9 M$).
It remains as future work to interpolate over the spin of the primary, however, the Chebyshev interpolation method appears to be a promising approach to tiling data from expensive gravitational self-force codes across the 4-dimensional generic Kerr parameter space.
This method could be further improved with the aid of a detailed of the study of the analytic structure of the GSF near the last stable orbit.

With an interpolated self-force model in hand, we computed inspirals using an action-angle formulation of the method of osculating geodesics (OG).
This approach is sketched in Ref.~\cite{Gair2011} and in we implement these equations of motion for generic (eccentric and inclined) inspirals about a Kerr black hole.
Our Mathematica implementation will be made publicly available on the Black Hole Perturbation Toolkit \cite{BHPToolkit}. 
For a binary with (small) mass ratio, $\mr$,  numerically solving the osculating geodesic equations takes minutes to hours due to the need to resolve the $\sim1/\epsilon$ oscillations in the orbital elements.
To overcome this, we follow Paper I \cite{NITs} and apply near-identity (averaging) transformations (NIT) which produce equations of motion that capture the correct long-term secular evolution of the binary but can also be rapidly numerically solved.

As a test of this formulation, we applied it to our eccentric, equatorial self-forced inspirals. 
We showed that our NIT'd quantities remain close to the original evolution variables throughout the inspiral at the expected order in the mass-ratio.
When the mass ratio is greater than $1:300$, we find the difference between year-long NIT and OG inspirals becomes significant for data analysis, reinforcing  the findings of Ref.~\cite{McCart2021}. Note, however, that a priori it is not known which (the NIT or OG inspiral) is closer to the true inspiral, since both are accurate to the same order in the mass-ratio.

With our efficient NIT model of eccentric, equatorial inspirals we explored the effects of the gravitational self force. 
We find that prograde inspirals around a rapidly rotating black hole generally experience an additional periastron advance on top of the periastron advance induced by geodesic motion. 
This is in contrast to the ``periastron retreat" experienced by retrograde inspirals and inspirals around non-rotating black holes \cite{LeTiec:2011bk}.

The NIT equations of motion make it convenient to compare inspirals both with and without post-adiabatic effects included and we confirmed that without post-adiabatic effects, the orbital phases of a typical EMRI will incur an error of order $\mathcal{O} (\mr^0)$. 
Moreover, by comparing inspirals under the influence of self-force models calculated in different gauges, we find that the resulting trajectories are gauge dependent. 
This difference due to gauge causes a de-phasing that is comparable in magnitude to not including any post-adiabatic effects. 
This suggests that in order to obtain gauge invariant post-adiabatic waveforms, one must also include second order self-force results. 
Second order self-force calculations are presently made using a two-timescale framework \cite{Warburton2021,Miller2021}. 
For the equations of motion, this framework is related to the NITs \cite{Kevorkian1987,Pound2021}, but more work is required in order to explicitly transform the two-timescale results into forcing terms that could be used in the framework presented in this paper.
Many of the averaging techniques developed in this work are also useful for the two-timescale approach \cite{Pound2021}.

For complete post-adiabatic waveforms, one would also need to include the spin of the secondary.
Inspirals incorporating the leading conservative spin induced effects around a non-rotating primary have been calculated \cite{Warburton2017} and the effect of (anti-)aligned secondary spin on the energy and angular momentum fluxes have recently been computed for eccentric orbits \cite{Skoupy:2021asz}. 
These effects can readily be incorporated into the NIT framework.

Another natural extension of our work is to non-equatorial orbital motion.
We already have results for spherical inspirals that will be the topic of an upcoming paper. 
After that we plan to tackle generic orbits, but there are two major barriers to this. 
The first is the larger parameter space which will make calculating the self-force extremely expensive \cite{Hughes2021}.
Our Chebyshev interpolation method should help to reduce the number of points in the parameter space where the self-force needs to be calculated.
The second barrier is the presence of orbital resonances \cite{Flanagan2012a,Flanagan2012b,vandeMeent2014a,Berry2016, Speri21}. 

Near these resonances the NITs break down and an alternative averaging procedure over the resonance timescale needs to be applied \cite{Pound2021}.  
While a lot is known about the effects of resonances on EMRI trajectories \cite{Flanagan2012a,Flanagan2012b,vandeMeent2014a,Berry2016, 
Mihaylov:2017qwn,Speri21}, rapidly computing inspiral trajectories while incorporating all resonant effects remains an open challenge.

Finally, we note that in this paper we use the leading-order quadrupole formula to generate the waveforms from the OG and NIT inspirals.
This is sufficient for our purposes where we wish to compare waveforms from OG and NIT inspirals but for LISA data analysis we will want to use relativistic waveform amplitudes.
These were recently efficiently interpolated in Ref.~\cite{Chua2021a} for Schwarzschild inspirals.
That work used orbit-averaged fluxes to drive the inspirals but it would be straightforward to use a NIT inspiral instead.
Once the waveform amplitudes have been interpolated for Kerr inspirals it could be combined immediately with the implementation presented in this work.

\section*{Acknowledgements}
PL acknowledges support from the Irish Research Council under grant GOIPG/2018/1978. 
NW acknowledges support from a Royal Society–Science Foundation Ireland Research Fellowship.
This publication has emanated from research conducted with the financial support of Science Foundation Ireland under Grant number 16/RS-URF/3428.
We thank Ian Hinder and Barry Wardell for the SimulationTools analysis package.
This work makes use of the Black Hole Perturbation Toolkit.

\appendix

\section{Geodesic motion}\label{apdx:geodesic}

We present here relations for quantities that were key to deriving our action-angle formulation for the method of osculating geodesics. 
Two of the radial roots the potential $V_r$ and one root of the polar potential $V_z$ are given in Eq.~\eqref{eq:primary_roots}.
The remaining roots are given by \cite{Fujita2009a}
\begin{subequations}\label{eq:secondary_roots}
	\begin{equation}
		r_3 = \frac{M}{1-\En^2} - \frac{r_1+r_2}{2} + \sqrt{\left( \frac{r_1+r_2}{2} -\frac{M}{1-\En^2} \right)^2 - \frac{a^2\Q}{r_1 r_2(1-\En^2)}}
	\end{equation}
	\begin{equation}
		r_4 =\frac{a^2\Q}{r_1 r_2 r_3(1-\En^2)},
	\end{equation}
	\begin{equation}
		z_+ = \sqrt{a^2(1-\En^2) + \frac{\Lz^2}{1-z_-^2}}.
	\end{equation}
\end{subequations}

Using action angles also has the advantage of providing analytic solutions for the $t$ and $\phi$ coordinates of the secondary, which take the form 
\begin{subequations} \label{eq:Extrinsic_Analytic_Solutions}
	\begin{gather}
		t(\lambda) = \Upsilon_{t} \lambda + t_r (q_r) + t_z (q_z)  \quad \text{and} \quad  \phi(\lambda) = \Upsilon_{\phi} \lambda + \phi_r (q_r) + \phi_z (q_z),    \tag{\theequation a-b}
	\end{gather}
\end{subequations}
where $\Upsilon_{t}$ and $\Upsilon_{\phi}$ are the Mino-time fundamental frequencies, $t_r$ and $\phi_r$ are periodic functions of $q_r$, and $t_z$ and $\phi_z$ are periodic functions of $q_z$.
The explicit expressions for these functions , and the Mino-time frequencies, can be found in Refs.~\cite{Fujita2009a,VandeMeent2020}, and are implemented in \texttt{KerrGeodesics} Mathematica package of the \textit{Black Hole Perturbation Toolkit} \cite{BHPToolkit}.

In this work we make use of analytic solutions to the geodesic equations written interms of the action angles for the orbital phases $\vec{q} = \{q_r,q_z\}$.
These were first derived in Ref.~\cite{Fujita2009a} and then presented in a simplified form in Ref.~\cite{VandeMeent2020}. 
The radial and polar solutions to the geodesic equations are given by
\begin{equation} \label{eq:r_analytic}
	r(q_r) = \frac{r_3 (r_1 - r_2)\sn^2\left(\frac{K(k_r)}{\pi} q_r | k_r\right) - r_2(r_1-r_3)}
	{(r_1 - r_2) \sn^2\left(\frac{K(k_r)}{\pi} q_r | k_r\right) - (r_1 - r_3)}
\end{equation}
and
\begin{equation} \label{eq:z_analytic}
	z(q_z) = z_-\sn\left(K(k_z)\frac{2(q_z+\frac{\pi}{2})}{\pi} | k_z\right),
\end{equation}
where $\sn$ is Jacobi elliptic sine function, $K$ is complete elliptic integral of the first kind, and
\begin{subequations}
	\begin{gather}
		k_r= \frac{(r_1 - r_2)(r_3-r_4)}{(r_1 - r_3)(r_2 - r_4)},  \quad \text{and} \quad  k_z= a^2 (1 - \En^2) \frac{z_-^2}{z_+^2}.    \tag{\theequation a-b}
	\end{gather}
\end{subequations}

\section{Evolution Equations for the Integrals of Motion} \label{EvolutionOfCoM}
Our goal for this section is to derive evolution equations for the integrals of motion $\vec{P} = \{p, e, x\}$. 
To do so, we must first consider how a different set of integrals of motion $\vec{\mathcal{P}} = \{\En, \Lz, \K \}$ evolve in terms of the covariant components of the particle's 4-acceleration $\{a_t, a_r, a_z, a_{\phi} \}$. 

Using the second osculating geodesic equation (\ref{eq:osculating_equations}b) along with definitions of $\En$ and $\Lz$ in Eq.~\eqref{eq:ELK}, we can obtain the evolution equations for $\En$ and $\Lz$:
\begin{subequations}
	\begin{gather}
		\frac{d \En}{d\lambda} = - \Sigma a_t,  \quad \text{and} \quad  \frac{d \Lz}{d\lambda} = \Sigma a_{\phi}.   \tag{\theequation a-b}
	\end{gather}
\end{subequations}
To find the the evolution of $\K$, we note that the contravariant components of the Killing tensor can be written as \cite{Gair2011}
\begin{equation}
	\K^{\alpha \beta} = 2 \Sigma l^{( \alpha} n^{\beta )} + r^2 g^{\alpha \beta},
\end{equation}
where $\vec{l}$ and $\vec{n}$ are null vectors with components
\begin{subequations}
	\begin{gather}
		\vec{l} = \frac{\varpi^2}{\Delta} \partial_t + \partial_r + \frac{a}{\Delta} \partial_{\phi} 
		\quad \text{and} \quad  
		\vec{n} = \frac{\varpi^2}{2 \Sigma} \partial_t -\frac{\Delta}{2\Sigma} \partial_r + \frac{a}{2\Sigma} \partial_{\phi}.   
		\tag{\theequation a-b}
	\end{gather}
	\end{subequations}
Taking the derivative of $\K$ from Eq.~(\ref{eq:ELK}c) with respect to proper time gives us
\begin{equation}
	\frac{d \K}{d \tau} = \K^{\alpha \beta} u_\alpha a_\beta. 
\end{equation}
Expanding this out explicitly while making use of the orthogonality condition, $g^{\alpha \beta}u_{\alpha}a_{\beta} = 0$ and converting to Mino time gives us:
\begin{equation}
	\frac{d \K}{d \lambda} =  \frac{d \En}{d \lambda} \frac{2}{\Delta} \left( \varpi^4 \En - a \varpi ^2 \Lz  \right) + \frac{d \Lz}{d \lambda} \frac{2}{\Delta}\left(a^2 \Lz - a \varpi^2 \En \right) - 2 \Sigma \Delta u_r a_r.
\end{equation}

Using the above equations, we can express how the roots $\{r_1, r_2, z_-\}$ evolve with Mino time by exploiting the same trick as in Appendix A.2 and A.3 of \cite{Gair2011}.
First, we note that, using the chain rule, we can express the rate of change of $r_1$ or $r_2$ as
\begin{equation} \label{eq:r12_chain_rule}
	\frac{d r_{1,2}}{d \lambda} = \frac{\partial r_{1,2}}{\partial \mathcal{P}_j} \frac{d \mathcal{P}_j}{d \lambda}.
\end{equation}
We then find expressions for $\partial r_{1,2}/\partial \mathcal{P}_j$ be differentiating $V_r(r)$ with respect to $\mathcal{P}_j$.
\begin{subequations}
	\begin{equation}
		\frac{\partial V_r}{\partial \mathcal{P}_j}\Bigr|_{\substack{r = r_1}} = (1- \En^2)(r_1-r_2)(r_1 -  r_3) (r_1 - r_4) \frac{\partial r_1}{\partial \mathcal{P}_j},
	\end{equation}
	\begin{equation}
		\frac{\partial V_r}{\partial \mathcal{P}_j}\Bigr|_{\substack{r = r_2}} = - (1- \En^2)(r_1-r_2)(r_2 -  r_3) (r_2 - r_4) \frac{\partial r_2}{\partial \mathcal{P}_j}. 
	\end{equation}
\end{subequations}
We then note that the coefficients proceeding $\partial r_{1,2}/\partial \mathcal{P}_j$ are also obtained by differentiating $V_r$ with respect to $r$ and then evaluating at $r_{1,2}$, i.e.,
\begin{equation}\label{eq:dr12dPi}
	\frac{\partial V_r}{\partial \mathcal{P}_j}\Bigr|_{\substack{r_{1,2}}} = - \kappa(r_{1,2}) \frac{\partial r_{1,2}}{\partial \mathcal{P}_j} ,
\end{equation}
where we have defined
\begin{equation}
	\kappa(r) \coloneqq \frac{d V_r}{d r} = 
	4 \En F(r) r -  2 r \Delta(r) - 2(r-M) (r^2 + \K),
\end{equation}
\begin{equation}
	F(r) \coloneqq \varpi(r)^2 \En - a \Lz.
\end{equation}
Combining equations (\ref{eq:r12_chain_rule}) and (\ref{eq:dr12dPi}) and using the appropriate definition of $V_r$ from equation (\ref{eq:Vr}) to calculate the partial derivatives gives us our evolution equations for $r_1$ and $r_2$:
\begin{equation}
	\frac{d r_{1,2}}{d \lambda} = -\frac{2 F(r_{1,2})}{\kappa(r_{1,2})} \left( \varpi(r_{1,2})^2 \frac{d \En}{d \lambda } - a \frac{d \Lz}{d\lambda} \right) + \frac{\Delta(r_{1,2})}{\kappa(r_{1,2})} \frac{d \K}{d \lambda}.
\end{equation}

We can use similar steps as above to find the evolution of $z_-$. Again, the chain rule tells us that the evolution of $z_-$ follows 
\begin{equation} \label{eq:zm_chain_rule}
	\frac{d z_{-}}{d \lambda} = \frac{\partial z_-}{\partial \mathcal{P}_j} \frac{d \mathcal{P}_j}{d \lambda}.
\end{equation}
We then use $\frac{\partial V_z}{\partial \mathcal{P}_j}\Bigr|_{\substack{z_-}}$ along with the second definition $V_z$ in equation (\ref{eq:Vz}) to find an expression for $\frac{\partial z_-}{\partial P_j}$
\begin{equation}\label{eq:zm_pds}
	\frac{\partial V_z}{\partial\mathcal{P}_j}\Bigr|_{\substack{z_-}} = -2 z_- (\beta z_-^2 -z_+^2) \frac{\partial z_-}{\partial \mathcal{P}_j}.
\end{equation}
However, using the first definition of $V_z$ in terms of $\{\En, \Lz, \Q \}$ gives us the following explicit expressions for $\frac{\partial V_z}{\partial \mathcal{P}_j}\Bigr|_{\substack{z_-}}$

\begin{subequations}\label{eq:Vz_pds}
	\begin{gather}
		\frac{\partial V_z}{\partial \En}\Bigr|_{\substack{z_-}} = 2 a^2 \En z_-^2 (1-z_-^2), 
		\quad
		\frac{\partial V_z}{\partial \Lz}\Bigr|_{\substack{z_-}} = -2 \Lz z_-^2,
		\quad \text{and} \quad  
		\frac{\partial V_z}{\partial \Q}\Bigr|_{\substack{z_-}} = 1-z_-^2.
		\tag{\theequation a-c}
	\end{gather}
\end{subequations}
Combining the results from equations (\ref{eq:zm_chain_rule}), (\ref{eq:zm_pds}) and (\ref{eq:Vz_pds}) gives us
\begin{equation}
	\frac{2z_- (z_+ - \beta z_-)}{(1-z_-^2)} \frac{d z_-}{d \lambda} = \frac{d \Q}{d \lambda} - 2 \Lz \left(\frac{z_-^2}{1-z_-^2} \right) \frac{d \Lz}{d \lambda} + 2 a^2 \En z_-^2 \frac{d \En}{d \lambda}.
\end{equation}
Since we have expressions for the evolution of $\{ \En, \Lz, \K \}$, we can derive and expression for the evolution of $\Q$ by taking the derivative of equation~\eqref{eq:Carter_Constant} with respect to $\lambda$:
\begin{equation}\label{eq:Q_Evolution}
	\frac{d \Q}{d \lambda} = \frac{d \K}{d \lambda} -2(\Lz - a \En)\left( \frac{d \Lz}{d \lambda} - a \frac{d \En}{d \lambda} \right).
\end{equation}
Combing these two results and simplifying yields our final expression for the evolution of $z_-$
\begin{equation}
	\frac{d z_-}{d \lambda} = \frac{1}{2 z_- (z_+^2 - \beta z_-^2)} \left( x^2 \frac{d \K}{d\lambda} - 2 \left(\Lz - a x^2 \En \right) \left(\frac{d \Lz}{d \lambda} - a x^2 \frac{d \En}{d \lambda} \right)  \right),
\end{equation}
where we've used Eq.~(\ref{eq:primary_roots}c) to tidy up the final expression.

Now that we know how $\{r_1,r_2,z_- \}$ evolve, determining the evolution of $\{p,e,x \}$ is straightforward since we can convert from one set to the other using the relations
\begin{subequations}\label{eq:roots_t_pex}
	\begin{gather}
		p = \frac{2 r_1 r_2}{M(r_1 + r_2)}, 
		\quad
		e = \frac{r_1 - r_2}{r_1 + r_2},
		\quad \text{and} \quad  
		x = \pm \sqrt{1-z_-^2}.
		\tag{\theequation a-c}
	\end{gather}
\end{subequations}
We can then take the derivative of these equations with respect to $\lambda$ and use the chain rule to obtain Eqs.~\eqref{eq:OrbitalElementEvolutionEqs}.

\section{Evolution of the Orbital Phases} \label{EvolutionOfPhases}

While it is straightforward to obtain Eq.~\eqref{eq:phase_eq_1} from the first osculating geodesic equation (\ref{eq:osculating_equations}a) this equation is difficult to evaluate numerically at the turning points of the orbit, i.e., when $\partial x_G^i/\partial q_i = 0$. 
As such, following the procedure described in Ref.~\cite{Gair2011}, we derive an equivalent evolution equation for the initial phases which is finite at the turning points.

We start by considering the definition of the geodesic potentials:
\begin{equation}\label{eq:potentials}
	V_i(x^i, \vec{P}) = \left( \frac{d x^i}{d \lambda} \right)^2.
\end{equation}
If we add together the derivative of both sides of this expression with respect to $P_j$ and then multiply both sides by $\dot{P_j}$, one obtains:
\begin{equation}
	\frac{\partial V_i}{\partial x^i} \left( \frac{\partial x^i}{\partial P_j} \dot{P_j}\right) + 
	\frac{\partial V_i}{\partial P_j} \dot{P_j}= 
	2\dot{x}^i \left( \frac{\partial \dot{x}^i}{\partial P_j} \dot{P_j}\right).
\end{equation}
Rearranging and plugging in Eq.~\eqref{eq:phase_eq_1} allows us to write
\begin{equation}\label{eq:phase_eq_2_step_1}
	\frac{\partial V}{\partial x^i} \frac{\partial x^i}{\partial q_i} \dot{q}_{i0} = \frac{\partial V_i}{\partial P_j} \dot{P_j} -2 \dot{x}^i \left(\frac{\partial \dot{x}^i}{\partial P_j} \dot{P_j} \right).
\end{equation}
We also note that taking the derivative of (\ref{eq:potentials}) with respect to Mino time $\lambda$ and rearranging yields
\begin{equation}
	 \frac{\partial V_i}{\partial P_j} \dot{P_j} = 2 \dot{x}^i \Ddot{x}^i -\frac{\partial V_i}{\partial x^i} \dot{x}^i. 
\end{equation}
Rearranging this and subbing into equation (\ref{eq:phase_eq_2_step_1}) and simplifying gives us
\begin{equation}\label{eq:phase_eq_2_step}
	\frac{\partial V}{\partial x_i} \frac{\partial x^i}{\partial q_i} \dot{q}_{i 0} = 2 \dot{x}^i \left( \left[ \Ddot{x}^i -\frac{1}{2} \frac{\partial V_i}{\partial x^i} \right] - \left(\frac{\partial \dot{x}^i}{\partial P_j} \dot{P_j} \right)\right).
\end{equation}

We note that the square bracket term will vanish for geodesics.
For perturbed orbits this means that this term will be proportional to the component of the four-acceleration $a^i$ scaled by a factor of $\Sigma^2$ to compensate for taking derivatives with respect to $\lambda$ instead of $\tau$. 
When evaluating this expression, we make use of the osculating condition $x^i(\lambda) = x_G^i(\lambda)$.
This leads is to the simplification
\begin{equation}
	\dot{x}^i = \dot{x}_G^i =  \frac{\partial x_G^i}{\partial q_i} \frac{d q_i}{d \lambda} = \frac{\partial x_G^i}{\partial q_i} \Upsilon_i.
\end{equation}
Combining these results with equation \eqref{eq:phase_eq_2_step} gives us our final expression for the evolution of the initial phases which is regular at the turning points, as expressed in Eq.~\ref{eq:phase_eq_2}.

\section{Self-force corrections to the periapsis advance around a spinning black hole}
\label{Periastron_Advance}

\begin{figure}
	\centering
	\includegraphics[width=\linewidth]{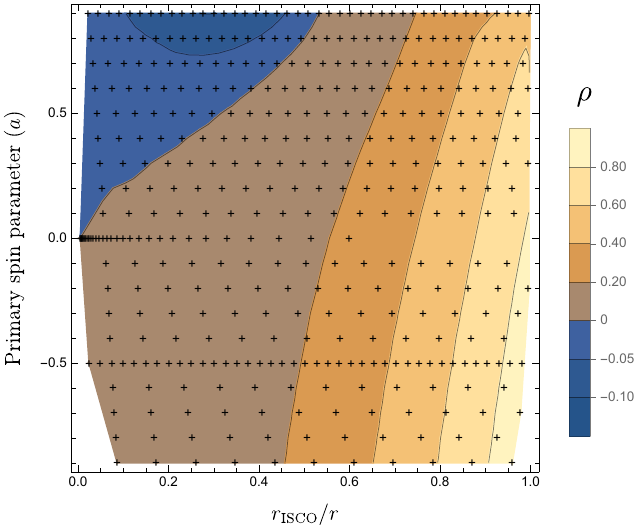}
	\caption{The linear in mass ratio correction, $\rho$, to the periapsis advance, $W$, as a function of distance from the innermost stable circular orbit (ISCO) $r_{\text{ISCO}}/r$ and the spin of the primary, $a$.
	 The contours show that $\rho$ grows larger as the radius of the inspiral approaches the ISCO. $\rho$ is positive for most of the parameter space, including for all retrograde orbits, implying that in these regions the self-force acts against the geodesic periapsis advance.
	 However, for all prograde orbits, there is a region (in blue) where this correction is negative meaning the self-force instead increases the rate of periapsis advance. 
	 The region where this occurs grows larger as $a$ increases
	 The black crosses mark the location of the underlying data from Ref.~\cite{VanDeMeent2017} used to calculate the contour plot.}
	\label{fig:PeriapsisAdvancePlot}
\end{figure}

The periapsis advance is a observable quantity that has been used to compare models of compact binary dynamics \cite{LeTiec:2011bk}. 
The effect of the gravitational self-force on this observable for quasi-circular EMRIs around a rotating primary was explored in Ref.~\cite{VanDeMeent2017}.
One important insight, that is present in the supplemental material of that work, is the effect that the spin of the primary and orbital radius has on the self-force correction to the rate of periapsis advance. 
For completeness, we highlight this result in this appendix.

For quasi-circular inspirals the relation between the dimensionless quantity $ W = \Omega_r^2 / \Omega_\phi^2 $ and $\Omega_\phi$ is an important benchmark for comparing between different calculational approaches to the two-body problem. \cite{LeTiec:2011bk,Tiec:2013twa,VanDeMeent2017}. The linear in mass ratio correction to the quantity is defined via
\begin{equation}
	W(\mr ; a,\Omega_\phi) = W(0; a,\Omega_\phi) + \mr \rho(a,\Omega_\phi) + \HOT{2},
\end{equation}
where $W(0; a ,\Omega_\phi)$ is the background value for the periapsis advance, and $\rho(a,\Omega_\phi)$ is the correction induced by the first-order gravitational self-force.

Fig.~\ref{fig:PeriapsisAdvancePlot} demonstrates how $\rho$ varies as a function of orbital radius $r$ and the spin of the primary $a$.
We plot the ratio $r_{\text{ISCO}}/r$, where $r_{\text{ISCO}}$ is the radius of the innermost stable circular orbit (ISCO). 
This ratio is convenient for plotting the results as goes from 1 at the  ISCO for all spin values and asymptotically approaches zero as $r$ grows large.
As one would expect, the plot demonstrates that this correction grows larger as the radius of the inspiral approaches the ISCO. 
This correction is positive for all retrograde orbits and in the strong field for prograde orbits. 
This means that self-force typically acts against the periapsis advance caused by the background geodesic motion, resulting in a reduction of the observed periapsis advance of the binary. 
However, for positive spins and at large radii, there is a region of the parameter space (in blue) where this correction is negative, meaning that the self-force increases the observed rate of periapsis advance compared to the background geodesic motion.
The larger the spin, the smaller the radii at which this effect occurs.
As such, this effect is most prominent for prograde orbits around rapidly rotating black holes. 

We find that the effect of the conservative self-force on the orbital phase for eccentric inspirals  is consistent with the sign of the self-force induced rate of periastron advance, $\rho$, in the quasi-circular limit -- see Sec.~\ref{section:AdbVsPA}.

\section*{References}
\bibliography{EccentricKerrNITsReferences_Revised}

\end{document}